\documentclass[acmsmall,screen]{acmart}
\AtBeginDocument{%
  \providecommand\BibTeX{{%
    \normalfont B\kern-0.5em{\scshape i\kern-0.25em b}\kern-0.8em\TeX}}}

\usepackage{csquotes}
\usepackage{graphicx}
\usepackage{tabularx}
\usepackage{array}
\usepackage{booktabs} 
\usepackage{colortbl} 
\usepackage{xcolor} 

\definecolor{lightgray}{rgb}{0.9,0.9,0.9} 
\definecolor{lightred}{rgb}{1.0,0.8,0.8} 
\definecolor{lightyellow}{rgb}{1.0,1.0,0.8} 
\definecolor{lightgreen}{rgb}{0.8,1.0,0.8} 
\definecolor{lightblue}{rgb}{0.8,0.8,1.0} 
\definecolor{lightpurple}{rgb}{0.9,0.8,1.0} 

\def\eg{\emph{e.g., }} 
\def\ie{\emph{i.e., }}

\setcopyright{acmcopyright}
\copyrightyear{2024}
\acmYear{2024}
\acmDOI{XXXXXXX.XXXXXXX}

\acmBooktitle{TiiS} 
\acmPrice{15.00}
\acmISBN{978-1-4503-XXXX-X/18/06}

\begin{document}

\title{ID.8: Co-Creating Visual Stories with Generative AI}



\author{Victor Nikhil Antony}
\affiliation{%
 \institution{Johns Hopkins University}
 \city{Baltimore}
 \state{Maryland}
 \country{U.S.A}}

\author{Chien-Ming Huang}
\affiliation{%
  \institution{Johns Hopkins University}
   \city{Baltimore}
 \state{Maryland}
 \country{U.S.A}}

\renewcommand{\shortauthors}{Antony and Huang}

\begin{abstract}
Storytelling is an integral part of human culture and significantly impacts cognitive and socio-emotional development and connection. Despite the importance of interactive visual storytelling, the process of creating such content requires specialized skills and is labor-intensive. This paper introduces ID.8, an open-source system designed for the co-creation of visual stories with generative AI. We focus on enabling an inclusive storytelling experience by simplifying the content creation process and allowing for customization. Our user evaluation confirms a generally positive user experience in domains such as enjoyment and exploration, while highlighting areas for improvement, particularly in immersiveness, alignment, and partnership between the user and the AI system. Overall, our findings indicate promising possibilities for empowering people to create visual stories with generative AI. 
This work contributes a novel content authoring system, ID.8, and insights into the challenges and potential of using generative AI for multimedia content creation. 

\end{abstract}

\ccsdesc[500]{Human-centered computing~Interaction design}
\ccsdesc[500]{Human-centered computing~Interactive systems and tools}

\keywords{Storytelling, Generative AI, Creativity}

 \begin{teaserfigure}
 \includegraphics[width=\textwidth]{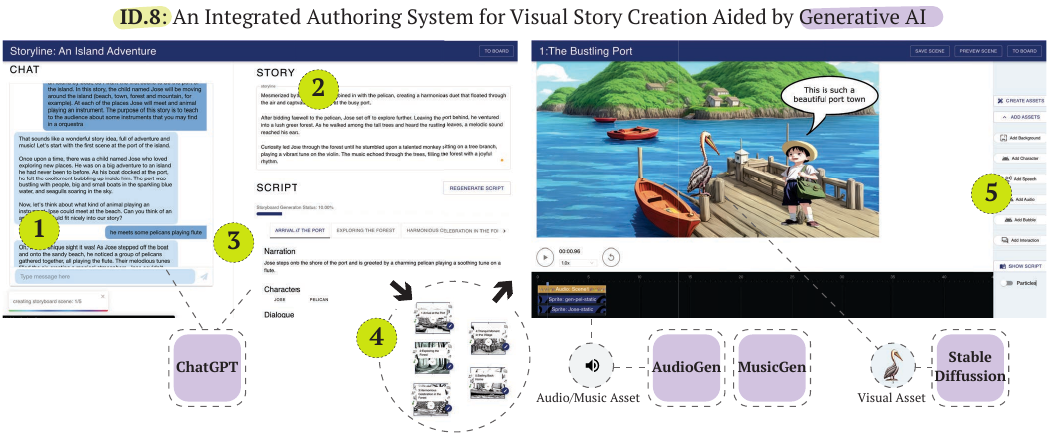}
 \caption{ID.8 features a multi-stage visual story authoring workflow facilitated by generative AI. (1) The story creation begins with users collaborating with ChatGPT to create a storyline and (2) manually editing of story content for finer edits. (3) Then, ID.8 automatically parses the story co-created with ChatGPT into a scene-by-scene script to be edited further by the user. (4) The story scenes from the script are automatically pre-populated and organized in the Storyboard and (5) edited in the Scene Editor where users use StableDiffusion, AudioGen, and MusicGen to generate story elements and synchronize story element on the canvas and the timeline.}
   \Description[A figure showing high-level overview workflow of ID8]{Figure 1 illustrates ID.8 and its multi-stage visual story authoring workflow facilitated by generative AI using annotated screenshots of the user interface. It begins by showing the Storyline creator module, starting with a chatbox with messages that depicts how the story creation in ID.8 begins with users collaborating with ChatGPT to create a storyline. To the right of this chatbox, the figure shows a text editing box filled with text of a box that depics how ID.8 allows for manual editing of story content for finer edits. Below the text editing box, the figure shows a section of the ID.8 interface with multiple tabs with each tab depicting a scene from the the screenplay that ID.8 automatically parses the story created with ChatGPT into a scene-by-scene script to be edited further by the user. Between the text editing box and the screenplay, the figure shows a loading bar depicting the progress of the automatic creation of scenes from the script that are used to populate the storyboard which is shown inside a circular area with dashed background below the screenplay box. There is an arrow that point from the screenplay to the storyboard and another arrow that points from the storyboard to the Scene Editor which is depicted to the right of the Storyline creator module. The Scene Editor shows a canvas depicting a scene from titled 'A bustling port' that shows an animated scenery of a port with a small rowboat docked and a young boy wearing a white shirt, blue shorts, a biege hat and a tote bag. There is a speech bubble on top of the boy with the text 'This is such a beatiful port'. There is an island in the distance with two peak and green fields everywhere. This scene was created using Stable Diffusion, a text-to-imag emode, and the figure points to a pelican that is on the port near the boat connecting it to a circular box with a enlarged version of the pelican and the term 'Stable Diffusion' written next to it. Below the Canvas, is the timeline which is depicted as a rectangular box which shows three elements of the scene, two characters (i.e. the boy and the pelican) and an audio element; each element is depicted as a rectangular box in its own row. Audio elements have a orange background. Visual Elements have a blue background. The Audio element has a dashed line connecting it to two boxes that have the text 'AudioGen' and 'MusicGen' respectively showing the names of the two audio generative models used in ID.8. Lastly the scene editor area also has a vertical toolbar to the right hand edge that shows the following buttons: 'Create Assets', 'Add Assets', and 'Show Script'; Under the 'Add Assets', there are button for 'Add Background', 'Add Character', 'Add Speech', 'Add Audio', 'Add Bubble', and 'Add Interaction'. Lastly there is a toggle button named 'Particles' on this toolbar.}
 \label{fig:teaser}
\end{teaserfigure}


\maketitle

\section{Introduction}
Storytelling is a defining aspect of the human experience that has been practiced through diverse media, such as written text, oral traditions, cave paintings, and more \cite{sugiyama2001narrative}. Visual stories are narratives that are augmented by various forms of media such as drawings, illustrations, animations, and videos, serving to enhance the overall storytelling experience \cite{caputo2003visual}. Visual stories tend to increase interest in and emotional engagement with the narrative even as they improve understanding and retention of the story's content \cite{mirkovski2019visual}; they can be an ideal medium for psycho-educational interventions \cite{bradlyn2003psychoeducational}, health communications \cite{van2022interactive}, language learning \cite{snyder1988foreign}, intergenerational bonding \cite{li2019story}, and creative self-expression \cite{rutta2019comic}. 

Despite the benefits and use cases of visual stories, their creation process remains a challenging, multifaceted task that unfolds via a sequence of essential steps---such as brainstorming to cultivate ideas, scripting to develop the narrative, storyboarding for visual planning, amassing the necessary media assets, piecing together the elements, and refining the content through editing---culminating in the distribution of the finalized story \cite{quah2022systematic}. The nature of multimodal asset creation and the technical demands of specialized software (\eg Adobe Creative Studio, Final Cut Pro) present a skill barrier that hampers both expert and novice creators from fully tapping into visual stories' creative potential. Lowering barriers to visual story authoring can enable the production of individualized and customized content that may lead to improved outcomes in varying use cases and amongst diverse populations.

Recent advances in generative AI have enabled the production of text \cite{brown2020language}, images \cite{oppenlaender2022creativity}, audio \cite{copet2023simple}, and videos \cite{li2023videogen} from user instructions; generative AI models hold the potential to help democratize the visual story authoring landscape. Human-AI co-creation is a paradigm wherein human users collaborate with AI with varying degrees of partnership to create a product \cite{rezwana2022identifying}. Several co-creative systems have explored how generative models may help users create storylines \cite{chung2022talebrush, osone2021buncho}, draw visuals \cite{lawton2023tool, zhang2022storydrawer} and compose music \cite{louie2020novice}. However, to the best of our knowledge, no systems exist, co-creative or otherwise, that enable the end-to-end authoring of visual stories. Leveraging generative AI to simplify the authoring process may help enable the quick and expressive creation of visual stories. Thus, we aim to explore the following research questions: \textbf{RQ1:} How can we integrate different generative AI models in an end-to-end visual story authoring system to support visual story creation process? \textbf{RQ2:} How do people co-create visual stories with a generative AI system?

Toward empowering users to effectively explore the creative possibilities of and quickly iterate over and generate visual stories, we built \textbf{ID.8} \textit{(ideate)}, an open-source, end-to-end authoring system for visual story creation with generative AI integrated into its workflow (see Fig. \ref{fig:teaser}). Our system enables users to collaborate with ChatGPT (a large language model) \cite{brown2020language} to co-write a script for the story, generate visual assets via Stable Diffusion (a text-to-image model) \cite{rombach2022high}, and generate audio assets with AudioGen (a text-to-audio model) \cite{kreuk2022audiogen} and MusicGen (a text-to-music model) \cite{copet2023simple}. Although the various content generation paradigms used in different stages of ID.8 may have been individually assesed in a co-creative setting, to the best of our knowledge, no studies thus far have evaluated a multi-modal co-creative experience that occurs during visual story authoring; Moreover, No open-source system exists that enables end-to-end generation of visual stories with multi-modal content (e.g. text content, visuals, audio effects). ID.8 embodies a ``human-in-control, AI-in-the-loop'' design framework to balance user autonomy and AI assistance. Our goal is to harmonize the control, agency, content safety, and human touch inherent to the manual story creation process with the creative variability and production efficiency of generative AI. Studying the co-creative process in a multi-modal domain such as visual story authoring can yield novel insights grounded in more realistic scenarios where Generative AI is poised to be used in the real world.

We conducted a two-phased evaluation of the ID.8 system to assess its usability and creative breadth. We found that ID.8 provided an enjoyable user experience and that users greatly appreciated the value of integrating generative AI in the visual story authoring workflow; moreover, users generated a wide variety of stories via our system, demonstrating its creative capabilities. 
Through this evaluation, we also gained a deeper understanding of the challenges faced by users while using a multimodal co-creative system, offering insights and design implications for future human-AI co-creative systems. 

This work makes three key contributions:
\begin{enumerate}
  \item We design, develop, and release ID.8: a novel, open-source, end-to-end system that enables visual story authoring via a unified interface and a human-AI co-creative workflow, aiming to lower the skill ceiling required for and enable the agile iteration of and broader creative expression in visual storytelling.
  \item Insights from two user evaluations highlight the current opportunities and challenges of multimodal content creation via state-of-the-art generative AI.
  \item We put forward a set of design guidelines for human-AI co-creative systems based on our experience and empirical evidence from evaluating ID.8.
\end{enumerate}

\section{Related Work}

Here, we explore the three areas of work related to the motivation and development of ID.8. Section \ref{sec:bg-story-authoring} highlights how conventional story authoring tools facilitate users in crafting visual stories but often place the burden of asset production on users. Section \ref{sec:bg-generative-ai} focuses on state-of-the-art generative AI models, exploring how advancements in this area hold the potential for democratizing content creation across modalities. Section \ref{sec:bg-co-creative-systems} discusses the potential of human-AI co-creative systems, which synergize the complementary capabilities of humans and AI in a collaborative creative process. Through ID.8, we seek to address the gap at the intersection of these areas by building a platform that leverages generative AI to enhance the visual story authoring experience in a co-creative setting and to help better understand the challenges of co-creation in a multi-modal setting.

\subsection{Story Authoring Tools}
\label{sec:bg-story-authoring}
Visual stories take many forms such as comics, animated videos, interactive informatics, etc. There have been attempts at simplifying the authoring process towards enabling expressive storytelling. PrivacyToons \cite{suh2022privacytoon} and DataToon \cite{kim2019datatoon} allow users to create comics by incorporating their own sketches and adjust the arrangement and panels through pen and touch interactions; PrivacyToons moreover supports the design and creative ideation yet remains limited in terms of viewer interaction. However, these systems only allow for creation of static stories.  ScrollyVis \cite{morth2022scrollyvis} is a web app for authoring, editing, and presenting data-driven scientific articles and allows for the integration of common sources such as images, text, and video; it supports multiple layers, dynamic narratives, however, is limited to creating interactive articles. Katika \cite{jahanlou2022katika} is an end-to-end system that simplifies the process of creating explainer motion graphics videos for amateurs; it provides a graphical user interface that allows users to create shots based on a script, add artworks and animation from a crowdsourced library, and edit the video using semi-automated transitions. These systems demonstrate the productivity improvements that story authoring tools provide to users however are still limited due to the burden of asset production being placed on the user. 

To bridge the gap that exists due to burden of assets generation being placed on end-users and limited interactivity, we designed ID.8 and sought to leverage generative AI to support each stage of the visual story authoring process to empower users to quickly explore the creative landscape and materialize their vision. Past work has evaluated asset generation using generative AI however they have been limited to a single modality (e.g. text, audio, images); With ID.8, we aim to enable an end-to-end authoring of visual stories by integrating various generative models in a workflow that unifies text, audio and video content thus allowing users to collaborate with generative AI in a more complex workflow; We open-source our system to enable the study of human-AI co-creation in this multi-modal domain.

\subsection{Generative AI and Content Generation} 
\label{sec:bg-generative-ai}
Breakthroughs in generative AI capabilities have opened up a plethora of possibilities across various domains specifically in supporting creative processes and democratizing content generation. Advancements in text generation capabilities using large language models (LLMs) (\eg GPT-3 \cite{brown2020language}, PaLM \cite{chowdhery2022palm}, LLaMA \cite{touvron2023llama}, etc) have been transformative for natural language processing tasks ranging from text summarization to code generation \cite{gozalo2023survey}; the versatility in executing a range of vastly different natural language tasks such as summarization and question-answering makes these LLMs particularly useful \cite{lewis2019bart}. In the context of storytelling, LLMs are beginning to be used for generating narrative structures, dialogues, and even full-length screenplays \cite{simon2022tattletale, ammanabrolu2020story}. While they should not replace human creativity entirely, they offer new avenues for brainstorming, initial drafting, and expediting the scriptwriting process \cite{knapp2023situating}. 

In the visual domain, text-to-image models (\eg Stable Diffusion \cite{rombach2022high}, Midjourney, Dall-E \cite{ramesh2022hierarchical}) have been employed to create photo-realistic paintings, animated illustrations, and even designs for fabrics or interiors \cite{zhang2023text, wu2021clothgan}. Moreover, image generation models are capable of accepting inputs beyond just text---for instance, sketches, images, masks, canny edges, pose points---to help guide the generation of images or to edit existing work thus potentially enabling expressive communication of intent from users \cite{voynov2023sketch, zhang2023adding}. Such technology can allow for rapid prototyping and experimentation, granting creatives, both amateurs and professionals, a new range of tools to express their artistic visions. 

Generative models have also been used for composing music that spans various genres, including classical, jazz, and electronic music and are even capable of generating more abstract audio effects such as battlefield backdrops. AI systems like OpenAI's Jukebox \cite{dhariwal2020jukebox} and recent models such as Meta's MusicGen \cite{copet2023simple} and AudioGen \cite{kreuk2022audiogen} are designed to either generate new compositions or assist composers in their creative processes. 

Despite the opportunities these state-of-the-art generative AI models represent, there is a lack of work on the incorporation of generative AI that supports content generation in various modalities into a unified authoring process and environment for visual story creation.
With ID.8, we aim to integrate the growing capabilities of generative models to empower individuals to create interactive multimodal content to support visual storytelling. 

\subsection{Human-AI Co-Creative Systems}
\label{sec:bg-co-creative-systems}
Human-AI co-creation refers to shared, collaborative creation of a creative product by human(s) and AI system(s) \cite{rezwana2022identifying}. Advancements in generative AI are opening new frontiers for human-AI collaboration leading to a wide array of work towards building and evaluating human-AI co-creative systems in a variety of domains, including music, design, dance, and writing. 

Several co-creative systems have explored various methods of supporting human-AI collaborative writing \cite{chung2022talebrush, rezwana2021creative, ippolito2022creative, swanson2021story}. WordCraft \cite{ippolito2022creative} is a text editor with a built-in AI writing assistant which showed the need for goal alignment and adaptability to users' expertise in co-creative systems. Talebrush \cite{chung2022talebrush} was designed to help users intuitively guide the narrative for co-created stories using a line sketching mechanism spotlighting the value of exploring more intuitive mechanisms for facilitating intent communication between the user and the AI. BunCho was built to support high-level and creative writing of stories for a text-based interactive game and was found to enhance the enjoyability of the writing process and to broaden the creative breadth of the stories \cite{osone2021buncho}. 

Co-creative systems have also been built to support visual artistic expression. Reframer \cite{lawton2023tool} presented a novel human-AI drawing interface that facilitates an iterative,  reflective and concurrent co-creation process while SmartPaint \cite{sun2019smartpaint} introduced a co-creative painting system that empowers amateur artists to turn rough sketches into complete paintings.  Virtual embodied conversational agents were incorporated into the co-creative drawing process in Creative PenPal \cite{rezwana2021creative} and found to potentially improve user engagement and the collaborative experience. Furthermore, co-creative AI systems have been built to enable creative drawing with children; StoryDrawer is a system that supports collaborative drawing and establishes how co-creative AI can help enhance creative outcomes and lead to higher engagement in the creative process \cite{zhang2022storydrawer}. AIStory, another co-creative system for children, simplified the prompting process for text-to-image models by replacing text input with icons which children could choose to generate visuals to accompany their narratives \cite{han2023design}. 

Beyond the text and visual domains, prior work has explored the human-AI co-creation in the music domain. COSMIC \cite{zhang2021cosmic} explored collaborative music generation with a chatbot whereas MoMusic \cite{bian2023momusic} presented a prototype collaborative Music Composition system that uses human motion as a key input signal yet again highlighting the unique domains and interaction modalities co-creative AI can enable. 

Together, these human-AI co-creative systems represent the nascence of a type of human-computer interaction that will become increasingly commonplace as the capabilities of generative models continue to expand and shape human creative work, highlighting the importance of designing effective human-AI interaction strategies to harness the power of generative AI in human creative processes. However, there is still an absence of human-AI co-creative systems that integrate generative AI of different output modalities in a single workflow and understanding of how users would collaborate with such a multi-modal system.
With ID.8, we seek to deepen our understanding of how people interact with multiple generative AI models (particularly LLMs, text-to-image and text-to-audio models) and incorporate their outputs in the creation of visual stories.

\section{ID.8: An Integrated Authoring System for Visual Story Creation}

ID.8 was developed following a design philosophy and criteria that emerged from a brainstorming session involving four members of the initial research team with backgrounds in human-computer interaction, cognitive science and computer science; this session was aimed at devising strategies to facilitate the user-friendly generation of visual stories through the integration of generative AI technologies.

\begin{table}[t]
\centering
\caption{System Design Criteria (DC) and Rationale for ID.8 }
\label{tab:design-criteria}
\begin{tabular}{p{0.90\textwidth}}
\midrule
\textbf{Design Criteria 1:} \colorbox{lightpurple}{Enable End-to-End Authoring}\\
\midrule
\midrule
Authoring visual stories is a complex, multi-stage process that includes brainstorming, scripting, storyboarding, gathering media assets, synchronizing individual elements, refining the narrative, and delivering the completed story \cite{quah2022systematic}. Given the lack of comprehensive tools in this space, ID.8 consolidates these disparate steps into a unified workflow, thereby empowering users to create visual stories more effortlessly. \\

\midrule
\textbf{Design Criteria 2:} \colorbox{lightpurple}{Facilitate Intuitive Narrative Assembly} \\
\midrule
\midrule
The complexity in visual story authoring arises from juggling diverse elements like narrative arcs, and spatio-temporal asset pacing, making it challenging to maintain a coherent view of the interactions of different narrative elements. ID.8 addresses this by helping users establish robust mental models of the narrative structure through clear interaction cues and visual markers among various story elements. \\

\midrule
\textbf{Design Criteria 3:} \colorbox{lightpurple}{Amplify Creative Exploration}\\
\midrule
\midrule
Generative models hold the potential for rapid exploration of a vast creative landscape, thereby amplifying individual creativity and imagination \cite{weisz2023toward}. ID.8 is designed to streamline the exploration and curation of these AI-generated artifacts, allowing users to seamlessly integrate these outputs into their creative vision. \\

\midrule

\textbf{Design Criteria 4:} \colorbox{lightpurple}{Safeguard Creative Autonomy} \\ \midrule
\midrule
Critical issues like the loss of user autonomy and the generation of potentially dangerous or unsafe creative outputs can manifest in co-creative systems, often due to AI decisions taking precedence over human input \cite{buschek2021nine}. ID.8 is designed to prioritize user control, empowering them to evaluate, select, and integrate AI-generated assets, thus effectively mitigating the risk of AI-driven decisions overriding user choices \cite{weisz2023toward}. \\
\midrule
\textbf{Design Criteria 5:} \colorbox{lightpurple}{Support Modularity and Extensions} \\ 
\midrule
\midrule
  With the rapid advancements in Generative AI technologies in mind, ID.8 is architected at the system design level for modularity and extensibility, allowing for the seamless integration of emerging state-of-the-art generative models. \\
\midrule
\end{tabular}
\end{table}

\textbf{Design Philosophy and Criteria.} ID.8 aims to balance user autonomy and AI assistance through a design philosophy centered on a ``human-in-control, AI-in-the-loop'' framework. We strive to combine the control, agency, safety and human touch intrinsic to manual story creation with the creative accessibility and efficiency gains offered by generative AI. To align this philosophy with our goal of simplifying visual story creation, we outline the our core design criteria (DC) in Table \ref{tab:design-criteria}.

\begin{figure}[h]
    \centering
    \includegraphics[width=\textwidth]{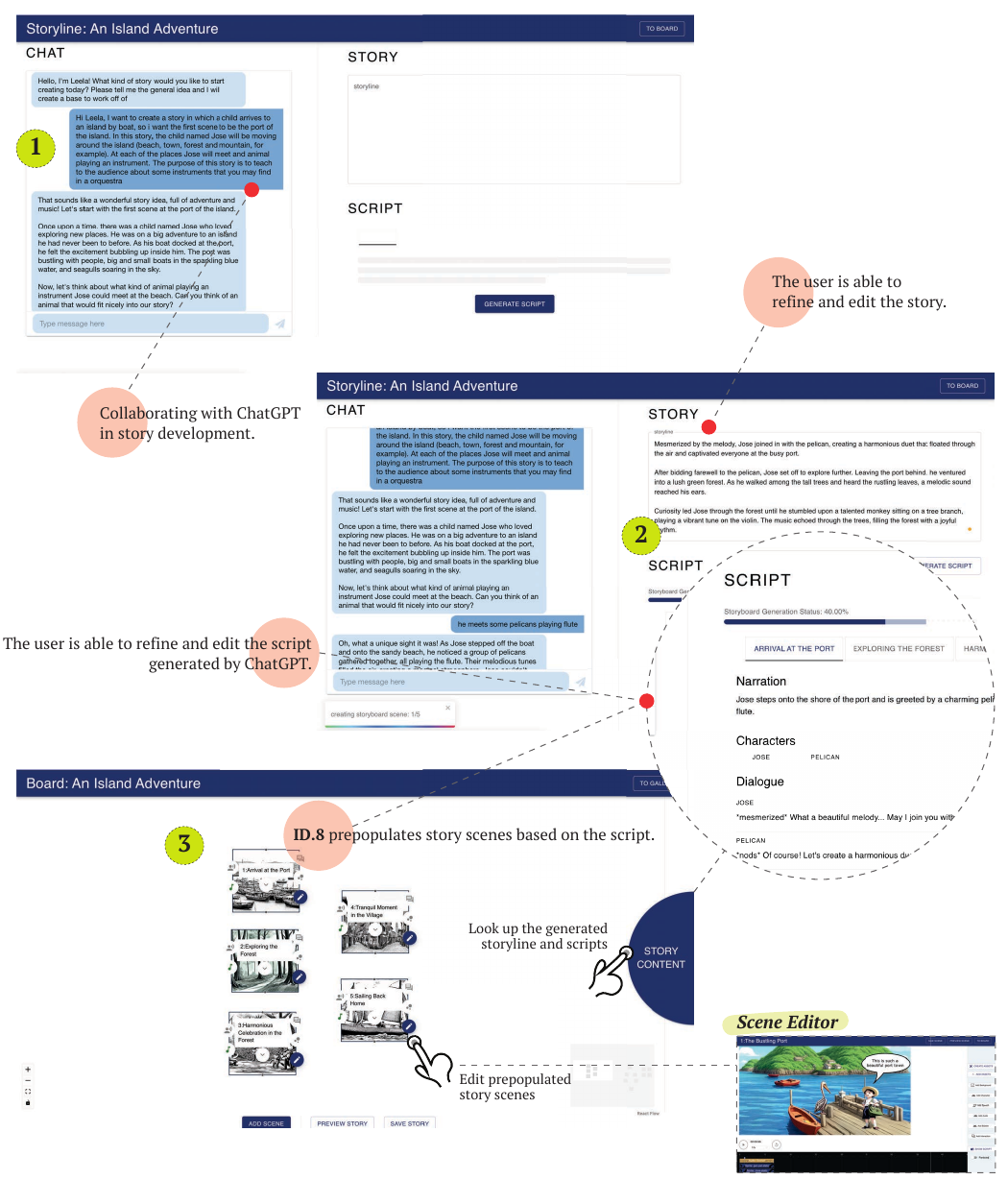}
    \caption[A figure showing the story text generation workflow and how it connects to the scene authoring]{ID.8 enables generation of a story (1) by collaborating with ChatGPT, and also allows the user (2) to manually edit the story and then (3) generates---using ChatGPT---a structured script and pre-populates the storyboard with scenes from the script.}
    \label{fig:sys-story-generation}
    \Description{Figure 2 illustrates the initial parts of ID.8 workflow starting from (1)-(2) the Storyline Creator Module and the (3) the Storyboard and Scene Editor. It is organized in three large sections and one small one sequentially layed out vertically. The first section is showing the Storyline creator module, starting with a chatbox with messages that depicts how the story creation in ID.8 begins with users collaborating with ChatGPT to create a storyline. To the right of this chatbox, the figure shows a text editing box with the heading 'Story' that is currently empty. Below this text box is the an area with a white bars and a blue button that says 'generate script'. The second section is showing the Storyline creator module, starting with a chatbox with messages that depicts how the story creation in ID.8 begins with users collaborating with ChatGPT to create a storyline. To the right of this chatbox, the figure shows a text editing box filled with text of a box that depics how ID.8 allows for manual editing of story content for finer edits. Below the text editing box, the figure shows a zoomed in view of the ID.8 screenplay interface with multiple tabs with each tab depicting a scene from the the screenplay that ID.8 automatically parses the story created with ChatGPT into a scene-by-scene script to be edited further by the user. The third section shows the storyboard section where it depicts 5 scenes as nodes each represented as a rectangle. It had three buttons on the bottom with the following text 'Add Scene', 'Preview Story' and 'Save Story'; It also had a semi-circular button 'Story Content' on the right hand side. The last section shows the Scene Editor which shows a canvas depicting a scene from titled 'A bustling port' that shows an animated scenery of a port with a small rowboat docked and a young boy wearing a white shirt, blue shorts, a biege hat and a tote bag. There is a speech bubble on top of the boy with the text 'This is such a beatiful port'. There is an island in the distance with two peak and green fields everywhere. This scene was created using Stable Diffusion, a text-to-image emode, and the figure points to a pelican that is on the port near the boat. Below the Canvas, is the timeline which is depicted as a rectangular box which shows three elements of the scene, two characters (i.e. the boy and the pelican) and an audio element; each element is depicted as a rectangular box in its own row. Audio elements have a orange background. Visual Elements have a blue background. Lastly the scene editor area also has a vertical toolbar to the right hand edge that shows the following buttons: 'Create Assets', 'Add Assets', and 'Show Script'; Under the 'Add Assets', there are button for 'Add Background', 'Add Character', 'Add Speech', 'Add Audio', 'Add Bubble', and 'Add Interaction'. Lastly there is a toggle button named 'Particles' on this toolbar.}
\end{figure}

\begin{figure}[t]
    \centering
    \includegraphics[width=\textwidth]{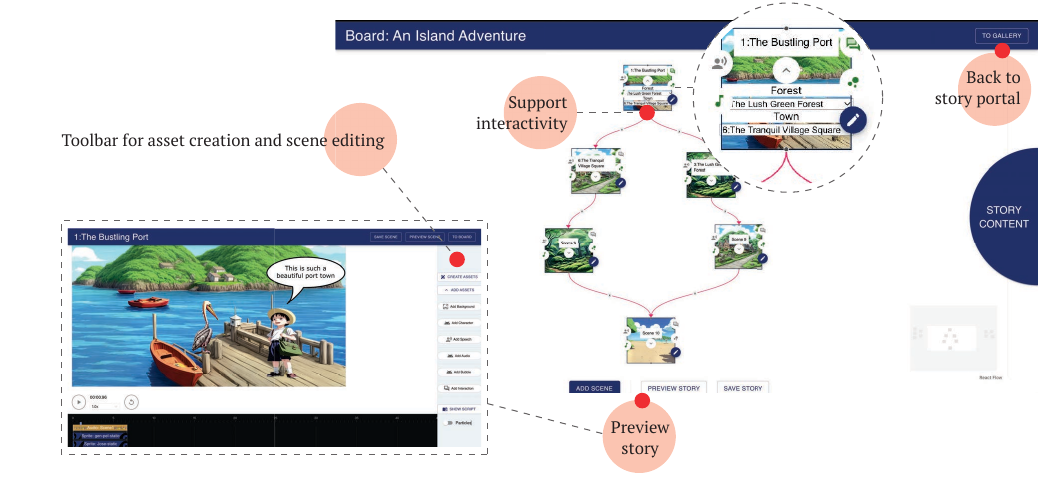}
    \caption{ID.8 Storyboard allows for organization of the story flow by linking scenes and specifying how story viewer inputs should impact the flow of the story. Users access the Scene Editor module by double-clicking a scene node. Users can also preview their story.}
    \label{fig:sys-storyboard}
    \Description[A figure showing the storyboard functionalities of ID8]{Figure 3 illustrates two boxes organized in a sequence horizontally. The first box depicts the Scene Editor which shows a canvas depicting a scene from titled 'A bustling port' that shows an animated scenery of a port with a small rowboat docked and a young boy wearing a white shirt, blue shorts, a biege hat and a tote bag. There is a speech bubble on top of the boy with the text 'This is such a beatiful port'. There is an island in the distance with two peak and green fields everywhere. This scene was created using Stable Diffusion, a text-to-image emode, and the figure points to a pelican that is on the port near the boat. Below the Canvas, is the timeline which is depicted as a rectangular box which shows three elements of the scene, two characters (i.e. the boy and the pelican) and an audio element; each element is depicted as a rectangular box in its own row. Audio elements have a orange background. Visual Elements have a blue background. Lastly the scene editor area also has a vertical toolbar to the right hand edge that shows the following buttons: 'Create Assets', 'Add Assets', and 'Show Script'; Under the 'Add Assets', there are button for 'Add Background', 'Add Character', 'Add Speech', 'Add Audio', 'Add Bubble', and 'Add Interaction'. Lastly there is a toggle button named 'Particles' on this toolbar. The second box depicts the storyboard section where it depicts 5 scenes as nodes each represented as a rectangle and connected to in the following links [1,2,5,4] and [1,3,6,4]. It had three buttons on the bottom with the following text 'Add Scene', 'Preview Story' and 'Save Story'; It also had a semi-circular button 'Story Content' on the right hand side. There is a button on the top right hand side 'To Gallery'. There is a zoomed in version of a scene node that depicts the scene Title and some text that depict the allowed user responses to possible questions with a drop down menu on available connected nodes.}
\end{figure}

\begin{figure}[h]
    \centering
    \includegraphics[width=\textwidth]{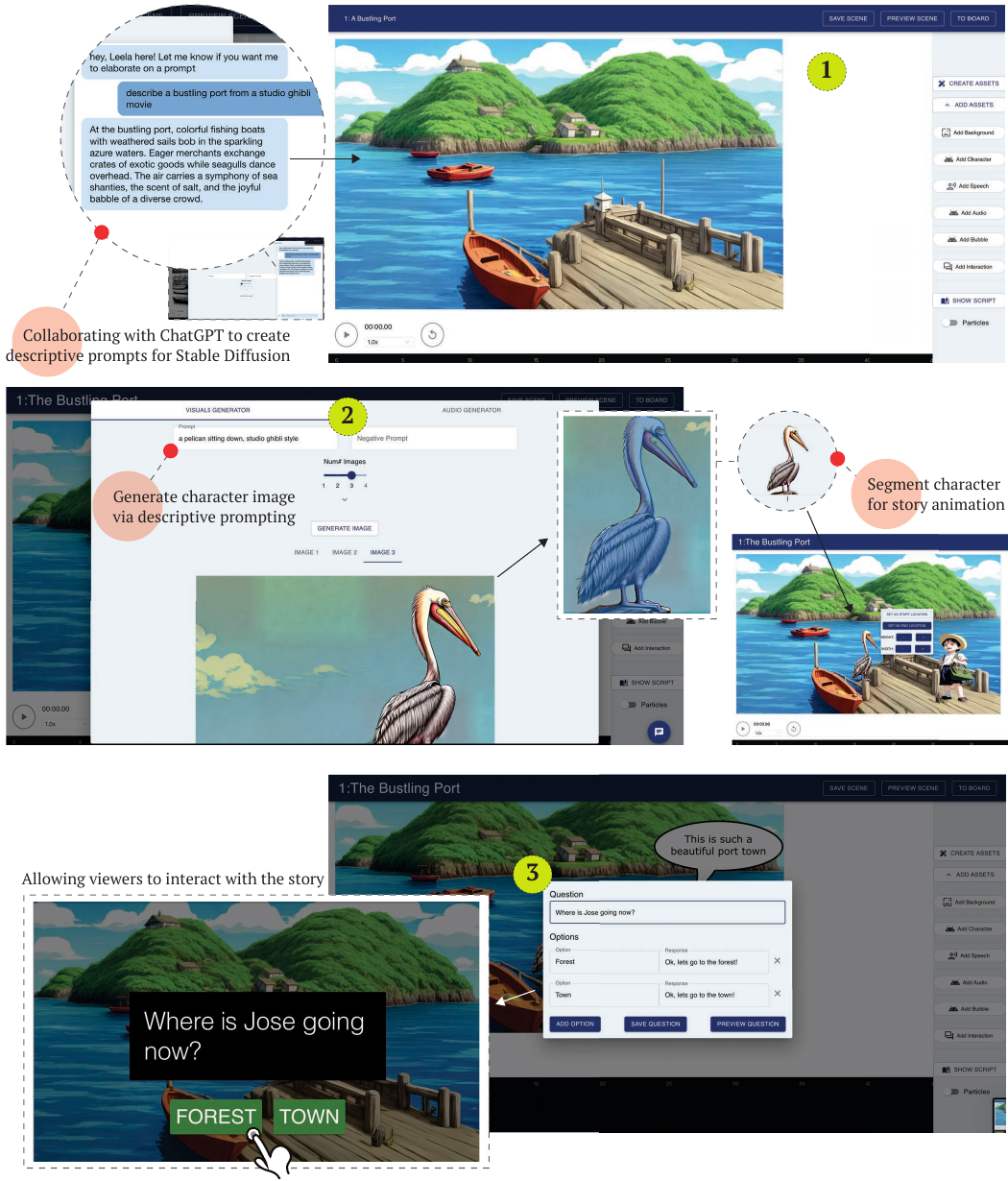}
    \caption{(1)The ID.8 Scene Editor enables creation of prompts for text-to-image/audio models in collaboration with ChatGPT; (2) For character generation, ID.8 empowers users to select parts of the generated output to be used in the story; (3) ID.8 provides a simple interface for adding interaction with viewer.}
    \label{fig:sys-esset-generation}
    \Description[A figure showing the asset generation and interaction specification workflow]{
Figure 4 illustrates four boxes organized in a sequence vertically. The first box depicts a zoomed in, focused view of a chatbox which shows a conversation on where ChatGPT helps the user create a prompt for a text-to-image model.The second box depicts the Scene Editor which shows a canvas depicting a scene from titled 'A bustling port' that shows an animated scenery of a port with a small rowboat docked and an island in the distance with two peak and green fields everywhere. Below the Canvas, is the timeline which is depicted as a rectangular box. Lastly the scene editor area also has a vertical toolbar to the right hand edge that shows the following buttons: 'Create Assets', 'Add Assets', and 'Show Script'; Under the 'Add Assets', there are button for 'Add Background', 'Add Character', 'Add Speech', 'Add Audio', 'Add Bubble', and 'Add Interaction'. Lastly there is a toggle button named 'Particles' on this toolbar. The third box depicts the asset generation section where there are two tabs: 'Visuals Generator' and 'Audio Generator' and the Visuals Generator is selected. Within the tab, there are two text fields for entering a prompt and negative prompt for the text-to-image model. The prompt text box has 'pelican sitting down, studio ghibli style' written on it. Below the text box, there is a slider to choose the number of sample images you want generated (1-4 range, discrete). Below the slider, there is a generate image button. Below this button, there is tabbed-interface with titles 'Image 1, 'Image 2' and 'Image 3', Image 1 is selected and an image of a pelican is shown. The pelican has an arrow pointing to a zoomed in verison of itself where only the pelican is highlighted in blue without the background selected. To the ridht of the main box, there is a smalled box show the scene editor again but with a a young boy wearing a white shirt, blue shorts, a biege hat and a tote bag along with a pelican on the canvas with the same background as before. There is an overlay interface on near the boy showing buttons for 'set start location', 'set end location', and buttons to increase and decrease height and width. The last section shows an interface to specify questions to be asked at the end of the scene; It has one text input for a question, then two columns and two rows on text input. each row has one text input for user choice and one text input for system response for said choice. there are three button in the bottom with text 'add response', 'save question', 'preview interaction'. Semi-overlapping with the section that was just described is a section with a black box rendered with text 'Where is Jose going now' and two green boxes with text 'Forest' and 'Town' shown below. Both these section have the scene editor in the background with half transparency as these interfaces are part of the scene editor section.}
\end{figure}

\subsection{System Overview} 
ID.8 features a multi-stage visual story authoring workflow facilitated by generative AI (see Fig. \ref{fig:teaser}). ID.8 currently enforces a linear, unidirectional creative process with distinct stages. Users initiate story creation with the \textit{Storyline Creator} module where they collaborate with ChatGPT (alias 'Leela') to create the story plot and manually perform finer edits of the co-crafted plot (See Fig. \ref{fig:sys-story-generation}-1). Then, ID.8 systematically converts the storyline into a scene-by-scene script to be edited further by the user. The story scenes from the script are automatically pre-populated and organized as individual nodes in the \textit{Storyboard} module. Users then edit each scene in the \textit{Scene Editor} where users use
StableDiffusion, AudioGen and MusicGen to generate story elements and synchronize various narrative elements on the canvas and the timeline. Users can watch the story as a whole or preview a single scene to experience the story as the viewer and adjust accordingly.

We developed ID.8 as an open-source\footnote{link to github: \url{https://github.com/vantony1/IDEATE.git}} ReactJS application to facilitate inter-platform accessibility. 
ID.8 uses REST API calls to remote servers (\ie OpenAI API\footnote{https://openai.com/product}, HuggingFace Spaces\footnote{https://huggingface.co/facebook/audiogen-medium}, StableDiffusionWeb API\footnote{https://stablediffusionapi.com/}) for interactions with the generative models, providing a flexible architecture that supports seamless model updates as new models being introduced.

\subsection{Storyline Creator}
The first stage of the ID.8 workflow is the Storyline Creator, a module that serves as a co-screenwriting environment. Within this module, ChatGPT powered chatbot---nicknamed 'Leela'---is pre-configured with an initial prompt (see Appendix \ref{apdx:system_prompt} for prompting details) to jointly craft narratives with the user (see Fig. \ref{fig:sys-story-generation}-1). In this setting, Leela poses questions, offers creative suggestions, and generates narrative components based on user input and feedback. This design enables users to focus on higher-level creative decisions, such as overarching plot design, character development, and thematic elements, while Leela supplements these efforts by producing nuanced narrative details and structural elements (DC1,2). Users retain complete creative control of the co-created narrative (see Fig. \ref{fig:sys-story-generation}-2) and are able to directly edit and refine the story generated with Leela (DC3). After the narrative is finalized, ID.8 compiles the co-crafted story into an editable, structured scene-by-scene script, complete with scene titles, narration, character lists, and dialogues---all generated via GPT-3.5 (see Appendix \ref{apdx:system_prompt} for prompting details). Moreover, ID.8 auto-generates corresponding scene nodes in the Storyboard, each featuring the scene's title and a monochromatic background visualized using Stable Diffusion (see Fig. \ref{fig:sys-story-generation}-3). This synthesis aims to bridge the narrative content with the visual scenes, thereby providing users with a foundation for further story construction (DC1).

\subsection{Storyboard} As a visual canvas (see Fig. \ref{fig:sys-storyboard}), the Storyboard module, supports an interactive, node-based visualization of the narrative (DC2). Users create and link individual nodes, each representing a specific scene, to create a cohesive story structure and flow. Metadata associated with each scene---such as titles, background settings, interactive components, and multimedia elements---are compactly displayed, aiding quick comprehension and navigation (DC2). The interface is designed to be intuitive, providing features like scene addition, deletion, and replication, as well as specifying a narrative starting point. To enrich the storytelling experience, the Storyboard module supports conditional narrative branching through its interaction components. This functionality aims not only to streamline the user's creative workflow but also to provide granular control over the narrative trajectory, ensuring that the Storyboard functions as both a planning tool and an interactive blueprint for the story (DC4).

\subsection{Scene Editor} The Scene Editor is a multi-faceted workspace that complements both the Storyline Creator and the Storyboard by enabling authoring and editing capabilities for individual scenes. Designed to enable quick creative exploration and iteration, it provides an array of tools to select, synchronize, and experiment with the audio-visual elements and the flow of a scene (DC1,2) (see Fig. \ref{fig:sys-esset-generation}).

\textbf{Canvas: The Visual Workspace.} At the center of the Scene Editor is the Canvas, serving as a visual workspace for real-time scene assembly. It provides an interactive interface where users can directly view and manipulate visual elements like characters and speech bubbles; an overlaid contextual menu appears when an element is selected, offering a range of customization options (\eg dimensions, start and end locations for an animation path). Dynamic background effects, mimicking natural phenomena like rain or snow, can also be set using the toolbar (see Fig. \ref{fig:sys-storyboard}) to enrich the visual storytelling experience.

\textbf{Timeline: Orchestrating Timing and Sequencing.} Situated below the canvas, the timeline offers a visual platform for coordinating the sequence and timing of scene components (\eg characters) (DC2). It aids in crafting a cohesive narrative flow by allowing users to chronologically synchronize and quickly adjust these elements, ensuring that the visual and auditory assets sync well with each other and the script.

\textbf{Asset Creator.} Anchored within the Scene Editor, the Asset Creator (see Fig. \ref{fig:sys-esset-generation}) serves as a platform for generating, selecting, and adapting both visual and audio assets. This tool, featuring a dual-tabbed interface comprising a `Visuals Generator' and an `Audio Generator', allows users to explore creative possibilities and materialize their storytelling vision.

\begin{itemize}

  \item \textbf{Visuals Generator.} Using Stable Diffusion \cite{rombach2022high}, a text-to-image model, the Visuals Generator supports image creation based on user-provided prompts. Users may save generated image(s) as background or extract parts to be saved as characters using Meta's 'Segment Anything' model \cite{kirillov2023segment} (see Fig. \ref{fig:sys-esset-generation}-2). Advanced controls for fine-tuning the model's output include negative prompts, a variable range for the number of generated images (1--4), denoising steps, and various modes like panorama and self-attention.
  
  \item \textbf{Audio Generator.} Focused on sound crafting, the Audio Generator is powered by Meta's AudioGen \cite{kreuk2022audiogen} and MusicGen \cite{copet2023simple} models, allowing for the creation of sound effects (\eg applause, wolf howls) and musical pieces (\eg lo-fi background tracks, classical carnatic compositions) based on user descriptions. Controls like duration (1--10 seconds), top-p and guidance scale afford more granular manipulation of audio generation.
\end{itemize}

The Asset Creator also integrates `Leela' (ChatGPT) to facilitate collaborative prompt authoring with the user (see Fig. \ref{fig:sys-esset-generation}-1) towards more effective visual and audio generation thus striving to help users realize their creative vision (see Appendix \ref{apdx:sidebar_prompt} for prompt). Moreover, it offers users preview functionalities and multiple output options for greater creative control (DC3). This setup ensures that the AI components play a supportive yet non-dominant role in the creative process (DC4).

\textbf{Speech Generator.} To enable narration and dialogue, Scene Editor includes a Speech Generation module powered by Google Cloud Text-to-Speech\footnote{https://cloud.google.com/text-to-speech/}. This feature allows users to select, customize and preview voice outputs as well as adjust parameters such as pitch and speed for creating more tailored voices. Users have the option to save their customized speech profiles to ensure consistent auditory experiences across multiple scenes (DC1).

\textbf{Viewer Interaction.} ID.8 aims to elevate viewer engagement by incorporating interactive elements that invite viewers' participation within the story (DC1). Towards accomplishing this, the Scene Editor provides an option to append interactive questions at the conclusion of individual scenes, aimed at capturing viewer input. Questions are both displayed on-screen and vocalized through Google Cloud Text-to-Speech, with a set of selectable responses also presented to the viewer. Depending on the selected response, specific auditory feedback is triggered, and the storyline may diverge accordingly. For scenes featuring interactive components, ID.8 enables the creation of conditional narrative branches. Users have the flexibility to dictate how viewer responses can influence subsequent scenes, thereby introducing a level of interactivity that has the potential to impact the story's direction (DC4). ID.8 provides a stand-alone story viewer platform that is able to play the generated stories. This viewer platform can be used by embodied agents (\eg virtual characters, social robots) to enrich the story viewing experience; for instance, a social robot can ``tell'' the story along with expressive movements to engage the viewer in domains such as education and therapy.




\section{Evaluation}
We conducted two studies to evaluate ID.8 and to understand how end-users interact and collaborate with generative AI to create visual stories. Study 1 sought to understand the usability of ID.8, the creative support by generative AI in the visual story authoring workflow, and whether users can effectively generate different story elements such as plot, background, characters, audio effects, etc. Study 2 aimed to gain a deeper understanding of the ID.8 creative breath and the co-creative user experience through an open-ended story generation task that participants engage in over a longer period time (\ie one week) outside of a controlled lab environment. We report details of the two study designs and study-specific findings in this section and describe the general lessons learned from the studies and design guidelines for human-AI co-creative systems in Section \ref{sec:lessons-learned}.

\subsection{Study Measures}
\label{sec:measures}
In both studies, we used two questionnaires to assess the usability and collaborative aspects of ID.8: the System Usability Scale (SUS) \cite{brooke1996sus} and the Mixed-Initiative Creativity Support Index (MICSI) \cite{lawton2023drawing}. The SUS is a 10-item, five-point Likert scale questionnaire specifically designed to provide a reliable metric for system usability. SUS Scores for positively worded items are calculated as (response $-$ 1) and for negatively worded items as (5 $-$ response); these scores are then summed and multiplied by 2.5 to yield a final SUS score ranging from 0 to 100. A SUS score above 70 is considered to signal good usability of a system \cite{bangor2009determining}.

MICSI is an 18-item scale created to assess human-machine creative collaboration \cite{lawton2023drawing} (see Appendix \ref{apdx:micsi_questionnaire}). 
It includes five sub-scales for \textit{Creativity Support}---namely Enjoyment, Exploration, Expressiveness, Immersion, and Results-Worth-Effort; each of these sub-scales is evaluated via a pair of seven-point Likert-scale questions, with the score derived from the mean response value for each question pair. 
Additionally, MICSI includes sub-scales for assessing \textit{Human-Machine Collaboration}, each evaluated through a separate seven-point Likert-scale question; these sub-scales cover Communication, Alignment, Agency, and Partnership. 
A score of 5 or above on these MICSI sub-scales signifies a positive user experience \cite{lawton2023drawing}. 
It is worth noting that the interpretation of the Agency sub-scale is different from the rest of sub-scales; lower scores suggest the system is perceived primarily as a tool, while higher scores imply it is seen more as a collaborative partner \cite{lawton2023drawing}. 
We also slightly adjusted the accompanying set of exploratory questions in MICSI, which focus on Contribution, Satisfaction, Surprise, and Novelty in relation to the final outcome; specifically, we replaced the term `sketch' with `story' to better align with the context of ID.8.

\subsection{Study 1: Evaluating Usability}

We conducted an in-lab system evaluation to understand the usability of ID.8, the creative support provided by generative AI in the visual story authoring workflow, and whether users can generate different story elements using ID.8. This evaluation was centered around a structured task that was designed to ensure that users interacted with various sub-components of ID.8.

\subsubsection{Study Procedure and Task}
After providing informed consent, participants filled out a pre-study questionnaire collecting demographics information (see Section \ref{sec:lab-participants}). Then, an experimenter presented a walk-through video demonstrating the system's functionalities to familiarize participants with ID.8. Following this, participants were allocated 75 minutes to compose a story of their choice, and to visualize three scenes from that story. Participants were instructed to utilize the Asset Creator module to generate at least one background visual, one piece of audio or music, and one static character; they were also asked to add interactive elements to their story. Besides these basic task requirements, participants were free to be as creative as they wanted. To better capture the co-creative process, we gathered interaction logs with the generative models. We also asked participants to keep notes on their experience and creation process during the session. After their allotted time, participants were asked to fill out a post-study questionnaire, which included the SUS and MICSI. Finally, a semi-structured interview was administered to explore in greater depth the participants' preferences, criticisms, and suggestions for ID.8 in the co-creative process. Participants were compensated with gift cards valued at the rate of \$15/hr. Appendix \ref{apdx:pre_study_survey}, \ref{apdx:semi_struct_interview} and \ref{apdx:micsi_questionnaire} present details on the pre-study and post-study questionnaires.


\subsubsection{Participants}
\label{sec:lab-participants}
We recruited 11 participants (6 female, 5 male) via online chat forums. Participants were aged 19 to 28 ($M=24.1,SD=2.10$) and had a wide range of backgrounds and expertise including robotics, law and innovation consultancy. Participants had a varied degree of familiarity with LLMs (very familiar=1, familiar=5, neutral=2, unfamiliar=2), familiarity with image/audio generation models (familiar=1, neutral=4, unfamiliar=1, very unfamiliar=4), and familiarity with creating visual content (very familiar=1, familiar=2, neutral=4, unfamiliar=2, very unfamiliar=1). We excluded one participant from analysis due to a data collection failure during their session.

\subsubsection{Results}

\begin{figure}[t]
    \centering
    \includegraphics[width=\textwidth]{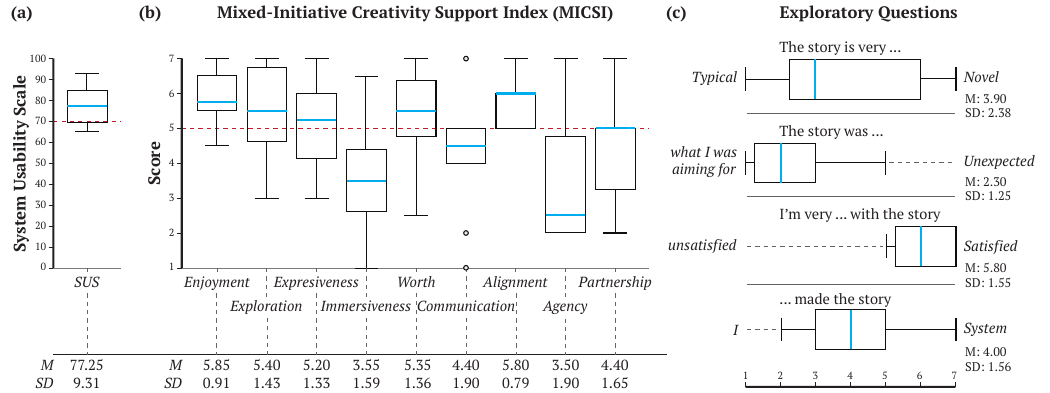}
    \caption{Results from Study 1: (a) SUS Scores, (b) MICSI Sub-Scale Scores, (c) Exploratory Question Responses.}
    \label{fig:lab-evaluation-results}
    \Description[A figure showing the quantitative results from Study 1]{Figure 5 illustrates 3 plots from Study 1: (1) SUS Scores, (2) MICSI Sub-Scale Scores, (3) Exploratory Question Responses. The first plot of SUS scores has one box plot and y axis range 0-100; The median is shown at 77.5 with a blue line, the range is 65-92.5, the 25th percentile mark is at 70 and the 75th percentile mark is at 84.4 -- Right below the plot the mean and standard deviations are given as 77.25 and 9.31. The second plot has 9 boxplots for each of the subscales of MISCI and y axis 1 to 7; The first box is for Enjoyment's median is shown at 5.75 with a blue line, the range is 4.5 to 7, the 25th percentile mark is at 5.5 and the 75th percentile mark is at 6.5 -- below the box plot the mean and standard deviations are given as 5.85 and 0.91; The second boxplot is for Exploration's median is shown at 5.75 with a blue line, the range is 3 to 7, the 25th percentile mark is at 4.625 and the 75th percentile mark is at 6.75 -- below the box plot the mean and standard deviations are given as 5.4 and 1.43; The third boxplot is for Expressiveness's median is shown at 5.25 with a blue line, the range is 3 to 7, the 25th percentile mark is at 4.125 and the 75th percentile mark is at 6 -- below the box plot the mean and standard deviations are given as 5.2 and 1.34; The fourth boxplot is for Immersiveness's median is shown at 3.5 with a blue line, the range is 1 to 7, the 25th percentile mark is at 2.625 and the 75th percentile mark is at 4.375-- below the box plot the mean and standard deviations are given as 3.55 and 1.59; The fifth boxplot is for Worth's median is shown at 5.5 with a blue line, the range is 2.5 to 7, the 25th percentile mark is at 4.75 and the 75th percentile mark is at 6.375-- below the box plot the mean and standard deviations are given as 5.35 and 1.34; The sixth boxplot is for Communication's median is shown at 4.5 with a blue line, the range is 1 to 7, the 25th percentile mark is at 4 and the 75th percentile mark is at 5-- below the box plot the mean and standard deviations are given as 4.4 and 1.90;  The seventh boxplot is for Alignment -- median is shown at 6 with a blue line, the range is 5 to 7, the 25th percentile mark is at 5 and the 75th percentile mark is at 6-- below the box plot the mean and standard deviations are given as 5.8 and 0.79;The eight boxplot is for Agency -- median is shown at 2.5 with a blue line, the range is 2 to 7, the 25th percentile mark is at 2 and the 75th percentile mark is at 4.75-- below the box plot the mean and standard deviations are given as 3.5 and 1.90; The ninth boxplot is for Partnership -- median is shown at 5 with a blue line, the range is 2 to 7, the 25th percentile mark is at 3.25 and the 75th percentile mark is at 5-- below the box plot the mean and standard deviations are given as 4.4 and 1.65; the third plot has box plot for responses to the exploratory questions; The first box plot is for the response comparing the statements: The story is very typical --- The story is very novel; it has mean 3.9, standard deviation 2.4, range 1-7, median is 3 and marked with a blue line, the 25th percentile is at 2.25 and the 75th percentile is at 6; The second box plot is for the response comparing the statements: The story was what I was aiming for --- The story outcome was unexpected; it has mean 2.3, standard deviation 1.25, range 1-5, median is 2 and marked with a blue line, the 25th percentile is at 1.25 and the 75th percentile is at 3; The third box plot is for the response comparing the statements: I'm very unsatisfied with the story --- I'm very satisfied with the story; it has mean 5.8, standard deviation 1.55, range 2-7, median is 6 and marked with a blue line, the 25th percentile is at 5.25 and the 75th percentile is at 7; The fourth box plot is for the response comparing the statements: I made the story --- The system made the story; it has mean 4, standard deviation 1.56, range 2-7, median is 4 and marked with a blue line, the 25th percentile is at 3 and the 75th percentile is at 6}
\end{figure}

We provide the quantitative results in Fig. \ref{fig:lab-evaluation-results} and report on key observations from the analysis of the survey data and transcriptions of the user interviews below. 

\begin{figure}[h]
    \centering
    \includegraphics[width=\textwidth]{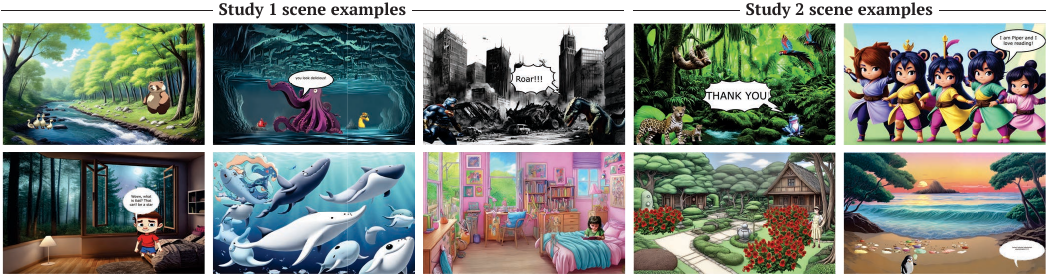}
    \caption{Scenes from stories generated by participants using ID.8 in Study 1 and Study 2.}
    \Description[A figure showing the scenes from stories generated by participants using ID.8 in Study 1 and Study 2.]{Figure 6 depicts 2 rows of 5 images showing scenes from stories generated by participants using ID.8 in Study 1 and Study 2. The top has a line running from one end to the other with a break between the 3rd and 4th images. The first part of the line has the caption Study 1 scene examples; The second part of the line has the caption Study 2 scene examples. The first image on the first row shows a bear standing next to a stream in a forest. The bear is brown and has a large head and body. It is standing on its hind legs and looking at the stream. The stream is flowing through the forest and is surrounded by trees and bushes. There are some ducks swimming in the stream. The ducks are white and brown and are paddling their feet to stay afloat. The overall scene is peaceful and serene. The next scene on this row shows a purple octopus standing in an underwater cave. The octopus is large and has eight tentacles. Its body is smooth and shiny, and its tentacles are long and thin. The octopus is facing the viewer and has its tentacles outstretched. There is a speech bubble coming from the octopus's mouth, and it says you look delicious. The cave behind the octopus is dark and rocky. There are some stalactites and stalagmites hanging from the ceiling. The floor of the cave is covered in dirt and rocks. The next scene in this row shows a comic book character and a dinosaur in a destroyed city. The comic book character is a man wearing a red and blue costume with a cape. He has a muscular build and a determined expression on his face. The dinosaur is a large, green creature with sharp teeth and claws. It is standing on its hind legs and roaring. The city in the background is in ruins. Buildings are collapsed and streets are littered with debris. There is smoke rising from the wreckage. The fourth image in the first row shows a collection of animals in a lush jungle scene. There is a sloth hanging from a tree branch, a jaguar stalking through the jungle, a frog sitting on a leaf, and a parrot perched on a branch. The animals are all different colors and textures, and they are arranged in a way that creates a sense of movement and energy. In the background of this scene, you can see the trees, vines, and flowers of the jungle. The sun is shining through the trees, creating a dappled effect on the ground. The last scene in the first row shows five animated princesses that are bubbly and wearing colorful clothes and holding handing standing in one line looking at us. The first princess has red hair and the rest have purplish hard. There is a speech bubble pointing to the fourth princess with the text 'I am piper and i love reading'. The first image in the second row shows a cartoon of a girl standing in a forest at night. She is wearing a blue dress and has long blonde hair. She is looking up at a rocket that is hovering in the air above her. The rocket is tall and slender, with a pointed nose and fins on the sides. It is emitting a bright light that illuminates the forest around it. There is also a cartoon boy wearing a red shit and blue shorts with a speech bubble pointing at him saying Wow, a spaceship. The moon can be seen through the canopy. The next image in this row shows a group of whales swimming in the ocean. The whales are all different sizes and colors, and they are swimming in a variety of directions. Some of the whales are jumping out of the water, and their fins and tails are glistening in the sunlight. The water in the background is blue and green. the third image in this row shows a cartoon of a little girl sitting on her pink bed in a pink bedroom, reading a book. She is wearing a nightgown and has her hair in pigtails. She is curled up under the covers, and the light from her bedside lamp casts a warm glow over her face. The girl is completely engrossed in her book, and she has a look of concentration on her face. the fourth image on the second row shows a garden with red flowers. The flowers are in full bloom and their petals are a deep, saturated red. The flowers are arranged in clusters and their stems are long and slender. The flowers are surrounded by green leaves and lush green grass. In the background of the painting, there is a wooden fence and a blue sky and some clouds. There is also a gardener in a white and yellow dress and she has short brunette hair. The last image on this row shows an animated penguin standing on a beach next to the ocean. The penguin is black and white with a long, slender body and flippers. It is standing on a rocky beach with sand and seashells scattered around. The ocean is in the background and it is a deep blue color. There are waves crashing on the shore and the sky is cloudy.
}
    \label{fig:story-other-examples}
\end{figure}

Participants found ID.8 to be easy to use (SUS score: $M=77.25,SD=9.31$) and recognized the value of the system in fast-tracking the visual story authoring. Moreover, participants were able to generate a wide range of scenes (see Fig. \ref{fig:story-other-examples}) with different artistic styles, characters and auditory elements using ID.8 and were satisfied with the authored story as indicated by the responses to the exploratory question regarding story satisfaction (see Fig. \ref{fig:lab-evaluation-results}-c) further highlighting the potential of generative AI in supporting the multi-modal creative process.

\begin{displayquote}
\textbf{P1}: \textit{``I think the image and the audio generation is quite interesting because yeah, if I would have done it myself, it would take a long time but using it is quite convenient for me and the chat with the LLM is quite useful. And I think the general user interface was easy to use.''}
\end{displayquote}

\begin{displayquote}
\textbf{P8}: \textit{``It was really helpful for quickly iterating on new ideas and exploring potential broad strokes of the story and for thinking about different ways you could represent a character or different events that could happen or the scenery or music. Relative to me trying to sketch those out myself, it was really efficient.''}
\end{displayquote}

The MICSI Enjoyment sub-scale scores ($M=5.85,SD=0.91$) and the feedback from the interviews show how ID.8 facilitated an enjoyable authoring experience. 


\begin{displayquote}
\textbf{P4}: \textit{``I feel like it's really cool. It's a really cool tool. Like I feel like I would use it again, if I had more time to play around with it because it's really fun.''}
\end{displayquote}

The overall positive user experience suggested by the Enjoyment, Exploration, Expressiveness, Worth, and Alignment sub-scales (Fig. \ref{fig:lab-evaluation-results}-b) indicates that ID.8 has reasonable realization of our design objectives of streamlining visual story authoring (DC1), enabling creative exploration (DC3), and maintaining user control (DC4).

 Participant responses to the exploratory question of who (System or I) made the story, suggest that users perceived their interaction with the system to be a balanced creative endeavor in terms of contribution. Yet, the Partnership score ($M=4.40,SD=1.65$) indicates a need for enhancing the sense of collaboration; this is further supported by the low Agency score ($M=3.50,SD=1.90$) indicating that the system was generally perceived not as a collaborative partner but rather as a tool. Similarly, the scores for Immersiveness ($M=3.55,SD=1.59$) and Communication ($M=4.40,SD=1.90$) lagged behind, signaling potential areas for improvement in the co-creative experience design. These lower scores suggest the need to refine the co-creative process to facilitate a more immersive user experience and to enhance mechanisms for effectively capturing user intentions. 

\subsection{Study 2: Evaluating Co-Creative Potential}
In order to better understand the creative width of ID.8 and to evaluate its human-AI co-creative experience in a less controlled setting, we conducted a longer-term, in-the-field study with participants who did not take part in Study 1 where we deployed our system on their personal devices and asked them to create a story over a week using ID.8. 

\subsubsection{Study Procedure and Task}
After providing informed consent, participants filled out a pre-study questionnaire collecting demographics information (see Section \ref{sec:lab-participants}). An experimenter deployed ID.8 using Docker images on participants' personal devices. Then, an experimenter presented the walk-through video shown in Study 1, demonstrating the system's functionalities to familiarize participants with ID.8. Participants were asked to use ID.8 to create a story of their choice and visualize it. We collected logs of their interactions with the generative models and asked them to keep a journal of their process and their experience. Once the participants finished creating their stories, we had them complete a post-study survey consisting of SUS and MISCI questions. Finally, a semi-structured interview was conducted to understand the participants' creative process along with their perceptions, preferences, criticisms, and suggestions for ID.8 and the co-creative process. Participants were compensated with a \$40 gift card. Appendix \ref{apdx:pre_study_survey}, \ref{apdx:semi_struct_interview} and \ref{apdx:micsi_questionnaire} present details on the pre-study and post-study questionnaires. 

\subsubsection{Participants}
We recruited 6 participants (3 females, 3 males) via online chat forums. Participants were aged 24 to 25 ($M=24.5,SD=0.5$) and had a range of backgrounds and expertise including visual arts, education, public health and electrical engineering. Participants had a range of familiarity with LLMs (neutral=3, unfamiliar=1, very familiar=2), familiarity with image/audio generation models (unfamiliar=1, very unfamiliar=5), and familiarity with creating visual content (very familiar=1, familiar=2, unfamiliar=2, very unfamiliar=2). 

\subsubsection{Results}

\begin{figure}[t]
    \centering
    \includegraphics[width=\textwidth]{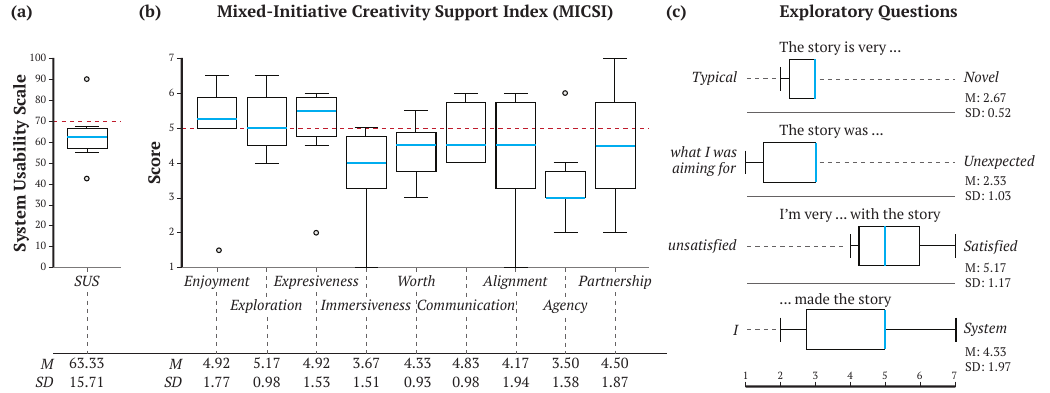}
    \caption{Results from Study 2: (a) SUS Scores, (b) MICSI Sub-Scale Scores, (c) MISCI Exploratory Scores.}
    \Description[A figure showing the quantitative results from Study 2]{
    Figure 7 illustrates 3 plots from Study 1: (1) SUS Scores, (2) MICSI Sub-Scale Scores, (3) Exploratory Question Responses. The first plot of SUS scores has one box plot and y-axis range 0-100; The median is shown at 62.5 with a blue line, the range is 42.5-90, the 25th percentile mark is at 56.875 and the 75th percentile mark is at 66.25 -- Right below the plot the mean and standard deviations are given as 63.33 and 15.71.  The second plot has 9 box plots for each of the subscales of MICSI and y-axis range 1 to 7; The first box is for Enjoyment—median is shown at 5.25 with a blue line, the range is 1.5 to 6.5, the 25th percentile mark is at 5 and the 75th percentile mark is at 5.875 -- below the box plot the mean and standard deviations are given as 4.917 and 1.772; The second box is for Exploration—median is shown at 5 with a blue line, the range is 4 to 6.5, the 25th percentile mark is at 4.5 and the 75th percentile mark is at 5.875 -- below the box plot the mean and standard deviations are given as 5.167 and 0.983; The third box is for Expressiveness—median is shown at 5.5 with a blue line, the range is 2 to 6, the 25th percentile mark is at 4.75 and the 75th percentile mark is at 5.875 -- below the box plot the mean and standard deviations are given as 4.917 and 1.53. the next box plot is for Immersiveness, the median is shown at 4 with a blue line, the range is 1-5, the 25th percentile mark is at 3.25 and the 75th percentile mark is at 4.75. Below the box plot, the mean and standard deviations are given as 3.667 and 1.506. the next box plot is for Worth, the median is shown at 4.5 with a blue line, the range is 3-5.5, the 25th percentile mark is at 3.75 and the 75th percentile mark is at 4.875. Below the box plot, the mean and standard deviations are given as 4.333 and 0.931. the next box plot is for Communication, the median is shown at 4.5 with a blue line, the range is 4-6, the 25th percentile mark is at 4 and the 75th percentile mark is at 5.75. Below the box plot, the mean and standard deviations are given as 4.833 and 0.983. the next box plot is for Alignment, the median is shown at 4.5 with a blue line, the range is 1-6, the 25th percentile mark is at 3.25 and the 75th percentile mark is at 5.75. Below the box plot, the mean and standard deviations are given as 4.167 and 1.941. the next box plot is for Agency, the median is shown at 3 with a blue line, the range is 2-6, the 25th percentile mark is at 3 and the 75th percentile mark is at 3.75. Below the box plot, the mean and standard deviations are given as 3.5 and 1.378. the next box plot is for Partnership, the median is shown at 4.5 with a blue line, the range is 2-7, the 25th percentile mark is at 3.25 and the 75th percentile mark is at 5.75. Below the box plot, the mean and standard deviations are given as 4.5 and 1.871.The third plot has box plots for responses to the exploratory questions; The first box plot is for the response comparing the statements: The story is very typical --- The story is very novel; it has mean 2.67, standard deviation 0.52, range 2-3, median is 3 and marked with a blue line, the 25th percentile is at 2.25 and the 75th percentile is at 3; The second box plot is for the response comparing the statements: The story was what I was aiming for --- The story outcome was unexpected; it has mean 2.33, standard deviation 1.03, range 1-3, median is 3 and marked with a blue line, the 25th percentile is at 1.5 and the 75th percentile is at 3; The third box plot is for the response comparing the statements: I'm very unsatisfied with the story --- I'm very satisfied with the story; it has mean 5.17, standard deviation 1.17, range 4-7, median is 5 and marked with a blue line, the 25th percentile is at 4.25 and the 75th percentile is at 5.75; The fourth box plot is for the response comparing the statements: I made the story --- The system made the story; it has mean 4.33, standard deviation 1.97, range 2-7, median is 5 and marked with a blue line, the 25th percentile is at 2.75 and the 75th percentile is at 5.}
    \label{fig:field-evaluation-results}
\end{figure}

We provide the quantitative results in Fig. \ref{fig:field-evaluation-results} and report on key observations from the analysis of the survey data and transcriptions of the user interviews below.

This evaluation yielded lower SUS scores ($M=63.33, SD=15.71$) and Enjoyment scores ($M=4.92, SD=1.84$) relative to Study 1; the factors resulting in these reduced scores remain uncertain---whether they stem from users encountering more bugs with extended system use, the complexity of initiating Docker containers via command line interfaces impacting usability, or a decrease in the novelty effect. Given the small sample size, it is difficult to draw robust conclusions about these outcomes. Further research is necessary to explore the factors influencing these results. Still, participants reported enjoying using the system in the interviews and were satisfied by the stories they were able to author using ID.8 as indicated by the exploratory question regarding story satisfaction (see Fig. \ref{fig:field-evaluation-results}-c).

\begin{displayquote}
\textbf{P14}: \textit{``Thank you for letting me use the tool. It was cool and now I have it on my laptop. So I'll keep using it.''}
\end{displayquote}

\begin{displayquote}
\textbf{P15}: \textit{``I just think that it was really cool and I would actually like to just use it on my own. I thought it was like, I feel like now that we did round one, if we did a round two would be way better, you know, like with what we could create.''}
\end{displayquote}

\begin{figure}[h]
    \centering
    \includegraphics[width=\textwidth]{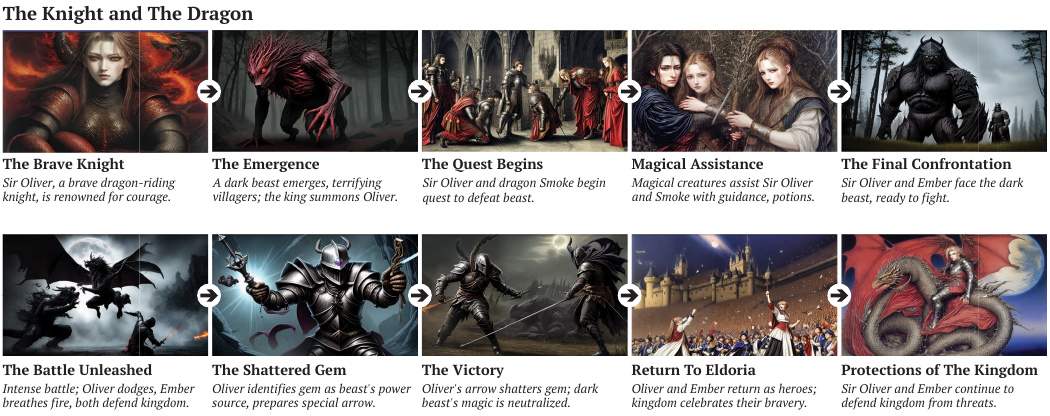}
    \caption{Scenes depicting the story titled 'The Knight and The Dragon' created by P13 using ID.8.}
    \Description[A figure showing scenes depicting the story titled 'The Knight and The Dragon' created by P13 using ID.8.]{Figure 8 scenes depicting the story titled 'The Knight and The Dragon' created by P13 using ID.8. The figure is arranged in two rows meant to be viewed and read left to right and top to bottom. The first scene is titled 'The Brave Knight' and has the caption 'Sir Oliver, a brave dragon-riding knight, is renowned for courage.' and shows Oliver in full plate armor standing in front of a large dragon. The knight is facing the viewer, and the dragon is behind him. The knight's armor and the dragon's scales are both red and brown colored. The background is an amalgamation of deep red undertones. The second scene is titled 'THE EMERGENCE' and has the caption 'A dark beast emerges, terrifying villagers - the king summons Oliver' and shows a terrifying monster with red eyes standing in a dark forest. The monster is tall and slender, with pale red-black skin and long, clawed fingers. It has a large head with sharp teeth and horns. Its eyes are glowing red, and it has a menacing expression on its face. The third scene has the title 'THE QUEST BEGINS' and the caption 'Sir Oliver and dragon Smoke begin quest to defeat beast.'. it shows a group of people in armor and red robes. Oliver is kneeling down in front of a kind who is standing in the center in armor. the scene is set in a medieval castle. The walls are made of stone, and there are torches burning in sconces on the walls. the fourth scene is titled MAGICAL ASSISTANCE and has caption Magical creatures assist Sir Oliver and Smoke with guidance, potions. It shows three mystical characters in a dark forest with concerned expressions on their face. the fifth scene is titled THE FINAL CONFRONTATION and has caption Sir Oliver and Ember face the dark beast, ready to fight. The scene shows shows a giant monster with horns and armor standing next to a knight in a forest. The monster is much taller than the knight, and its body is covered in scales. It has two large horns on its head, and its eyes glow red. The knight is wearing a suit of armor and wielding a sword. The forest in the background is dark and dense. The trees are tall and their branches are intertwined. The ground is covered in leaves and moss. The sixth scene which starts on the next row is titled THE BATTLE UNLEASHED and has the caption Intense battle, Oliver dodges, Ember breathes fire, both defend kingdom. It shows a knight and a dragon fighting a dark beast. The knight is wearing a suit of armor and wielding a sword. The dragon is large and has a long, scaly body. The beast is snarling and baring its teeth. the seventh scene is titled THE SHATTERED GEM and has caption Oliver identifies gem as beast's power source, prepares special arrow. It shows shows a knight in armor standing in a cave. The knight is holding two artefacts one with a light beaming out of it. The knight's armor is made of metal and covers the knight's entire body. The knight's helmet is also made of metal and covers the knight's head. The cave is dark and damp. The walls of the cave are made of rock. The next scene is titled THE VICTORY and has caption Oliver's arrow shatters gem; dark beast's magic is neutralized. It shows two knights standing next to each other. They are both wearing suits of armor and wielding swords. The knights are facing each other, and their swords are ready. The background is dark and gloomy and the moon is peering through night sky covered in musky clouds. The ninth scene is titled RETURN TO ELDORIA and captioned Oliver and Ember return as heroes; kingdom celebrates their bravery. it shows a massive crowd celebrating near a huge palace. Two women can be seen in medival clothes rejoicing. The last scene is titled PROTECTORS OF THE KINGDOM and captioned: Sir Oliver and Ember continue to defend kingdom from threats. It shows Oliver riding a dragon. The knight is wearing armor and is drapped in a red robe and has long, flowing hair. He is sitting on the dragon's back who  is large and green, with scales that shimmer in the moonlight. It has large red wings that are spread wide.
}
    \label{fig:story-examples}
\end{figure}

This study resulted in a collection of artistically diverse stories each about 5 minutes in length (see Fig. \ref{fig:story-other-examples}) spotlighting the creative width of ID.8\footnote{link to google drive folder with sample stories from study 2: \url{https://tinyurl.com/y3hjswhc}}. \eg P13 created an visually engaging story, see Fig. \ref{fig:story-examples}, with captivating audio elements and dramatic narration using ID.8 with a dark, grunge atmosphere with fantastical theme. The creative diversity of the authored stories and the Exploration ($M=5.17, SD=0.98$) score, see Fig. \ref{fig:field-evaluation-results}-b, further establish that ID.8 supports creative exploration reasonably well (DC3). 

The Alignment ($M=4.17,SD=1.94$) and Communication ($M=4.83,SD=0.98$) sub-scale scores indicate a need to  enable users to effectively communicate their intent with the co-creative system.  

This evaluation supports the findings of Study 1 regarding the need to enhanced sense of collaboration between the user and ID.8; participants again felt that their contributions were equally matched by the system (see Fig. \ref{fig:field-evaluation-results}-c) but the Partnership ($M=4.50,SD=1.87$) and Agency ($M=3.50,SD=1.38$) scores spotlight the need for improving the collaborative experience towards ID.8 being percieved as a collaborator than just a tool. The low Immersiveness ($M=3.67, SD=1.51$) score further underscore the need to facilitate a more immersive co-creative experience. 

Users reported spending 7--8 hours exploring the system and producing their stories. It was consistently reported that creating the first few scenes took most of the time; however, as they got used to the system, the creation process became faster. The manifestation of the ID.8's learning curve and discovery of shortcomings in the co-creative process in Study 2's less constrained creative setting highlights the value of conducting an open-ended, longer-term, in-the-field evaluation to gain a well-rounded perspective on a system's usability.

\begin{displayquote}
\textbf{P12}: \textit{``By the end, I feel like I was more used to knowing what to do and each scene definitely the time spent became like shorter and shorter.''}
\end{displayquote}

\section{Lessons Learned and Design Guidelines}
\label{sec:lessons-learned}

As generative AI increasingly influences the nature of creative work and content production across domains, it has been suggested that the prominent paradigm of human-AI interaction will be co-creation centered around communicating with generative models, with users supplying higher-level directions to complete complex downstream tasks \cite{liu2022design, oppenlaender2023prompting}. There are key design challenges in empowering users to effectively communicate their intent, which will in turn enhance the immersiveness of the co-creation process, strengthen the sense of partnership between the AI and the user, and promote the perception of the AI as a collaborator. Our exploratory findings of participants' experiences with ID.8 can serve as a proxy for the kinds of multi-agent, multimodal human-AI interactions that are predicted to become the norm in the future of creative work.  Here, we provide the key lessons learned from our evaluation of ID.8 regarding the potential and challenges of using generative AI and present a set of synthesized design guidelines (see Table \ref{tab:design-guidelines-cocreatives}) to inform the development of the interaction and workflow architectures of near-future human-AI co-creative systems. These design guidelines emerged from an inductive analysis of the transcriptions of the user interviews, system logs of user interaction with ID.8 and post-study survey data from our two studies.

\begin{table}[hb]
    \centering
    \caption{Design Guidelines for Human-AI Co-Creative Systems}
    \label{tab:design-guidelines-cocreatives}
    \begin{tabular}{p{0.24\linewidth}p{0.72\linewidth}} 
        \multicolumn{2}{l}{\textbf{Challenge:} Users' Mental Models of Generative AI Are Weak} \\
                
        \midrule
        \midrule
        \textbf{Design Guideline 1} & \colorbox{lightyellow}{Offer predesigned prompt templates with defined fields}to streamline the input process and facilitate user intent communication with generative models. \\
        \hline
        \textbf{Design Guideline 2} & \colorbox{lightyellow}{Provide a library of example outputs}alongside the prompts that generated them to enhance users' understanding of how to effectively guide generative models. \\
        \hline
        \textbf{Design Guideline 3} & \colorbox{lightyellow}{Leverage LLMs to assist users in crafting prompts}that effectively communicate their creative vision to generative models. \\
        \hline
        \textbf{Design Guideline 4} & \colorbox{lightyellow}{Use intuitive and semantically accurate descriptors or proxies for} \colorbox{lightyellow}{advanced model parameter controls}to improve user understanding and increase usage. \\
        \midrule
        \\
        \multicolumn{2}{l}{\textbf{Challenge:} Generative AI Outputs Biased and Harmful Content} \\
        \midrule
        \midrule
        \textbf{Design Guideline 5} &\colorbox{lightyellow}{Integrate safeguard measures such as trigger warnings and automatic} \colorbox{lightyellow}{content filters} against harmful generative outputs to ensure the emotional well-being of users. \\
        \midrule
        \\
        \multicolumn{2}{l}{\textbf{Challenge:} Strong Human-AI Partnership Requires Intuitive Communication} \\
        \midrule
        \midrule
        \textbf{Design Guideline 6} & \colorbox{lightyellow}{Accept multiple input modalities}to enable intuitive communication of intent. \\
        \hline
        \textbf{Design Guideline 7} & \colorbox{lightyellow}{Support iterative co-creation}through dynamic feedback loops. \\
        \hline
                \textbf{Design Guideline 8} & \colorbox{lightyellow}{Provide parallel processing}capabilities to facilitate a more dynamic, active co-creation experience. \\

        \midrule
        \\
                \multicolumn{2}{l}{\textbf{Challenge:} Co-Creative AI Needs a Unified Identity and an Adaptive Workflow} \\
        \midrule
        \midrule
        \textbf{Design Guideline 9} & \colorbox{lightyellow}{Establish a centralized AI identity}within the system to streamline user experience, enhance the sense of partnership, and increase immersion. \\
        \hline
        \textbf{Design Guideline 10} & \colorbox{lightyellow}{Adapt to individual creative workflows}to encourage the exploration of a variety of creative directions. \\
    \end{tabular}
\end{table}

\subsection{Users' Mental Models of Generative AI Are Weak}

While participants' experiences with ID.8 revealed both strengths and weaknesses of generative AI's capabilities, a significant barrier to effective interaction lies in users' limited mental models of how these systems work, particularly in the area of prompt engineering for content generation.

\subsubsection{Addressing Prompting Difficulties with Prompt Templates}
 Participants appreciated leveraging generative models to create multimedia assets for their stories; specifically, they valued the ability to quickly explore a wide range of creative possibilities and the workflow's efficiency as compared to manual asset creation. However, participants felt limited in their ability to precisely describe the specific assets they wanted to generate. For instance, they encountered difficulties in 1) avoiding generation certain unwanted content (\eg unnatural anatomy); 2) generating the same character with varying poses, clothing, and facial expressions; and 3) maintaining stylistic consistency across different generated assets. While some participants tried using specific artists' names in their prompts to achieve stylistic consistency, they reported instances where the generated image would disproportionately reflect the likeness or subject matter focus of a given artist. The struggle to effectively guide the generative models to produce desired outputs is also reflected in the suboptimal MISCI subscale scores for Communcation from both studies and the low Expressiveness and Alignment MISCI sub-scale scores from Study 2 (see Fig. \ref{fig:lab-evaluation-results} and Fig. \ref{fig:field-evaluation-results}).

\begin{displayquote}
    \textbf{P11}: \textit{``Visuals generation, like it was gorgeous in a lot of the cases, but it wouldn't get what I wanted or like it would give me weird things like two-headed animals.''}
\end{displayquote}

Our observations indicate that the process of prompting generative models is not intuitive enough to the average person; there is a stark need to simplify the art of prompt engineering \cite{zamfirescu2023johnny}. Providing prompt templates (\eg\textit{[Medium][Subject][Artist(s)][Details][Image repository support]} \cite{oppenlaender2022taxonomy}) with a wide set of prompt modifiers (\eg style specifiers, quality boosters \cite{oppenlaender2022taxonomy, pavlichenko2023best}) may be an effective approach toward building stronger user mental models of how to communicate intent and creative vision to generative AI models. While research exists on prompting techniques for text-to-image and LLMs, there is a significant gap in how to prompt generative models for other modalities (\eg audio, video, speech); therefore, future work should focus on creating modality-specific prompt templates to fully leverage the co-creative potential of generative AI in multi-modal settings.

\textbf{Design Guideline}: \textit{Offer pre-designed prompt templates with defined fields (e.g., [Medium][Subject] [Artist(s)][Details]) to streamline the input process and facilitate user intent communication with generative models.} 

\subsubsection{Building Intuition on Model Capabilities Using a Prompt-Output Pair Library}

Generative models have very impressive capabilities, but their outputs are greatly dependent on the prompts that they are provided with. We noticed that the difficulties participants experienced while prompting the models resulted in misunderstanding what the models are capable of generating. For instance, participants reported having trouble generating specific animals (\eg narwals, dragons) and maintaining stylistic consistency across generations; similarly, even though participants were generally successful in prompting Leela (ChatGPT), some desired more emphasis on character development, the inclusion of thematic elements such as a ``moral of the story'' and the use of simpler language for younger target audiences. Interestingly, ChatGPT is capable of making these changes when asked further, spotlighting a lack of coherent understanding of the model's capabilities. 

\begin{displayquote}
    \textbf{P12}: \textit{``For me, it was really hard making consistent images. And I felt like it was really hard creating like a cohesive image group. like the cave scene literally looked like somebody took a picture of a cave and the next scene has some anime girlies fighting a thing. It just feels super shocking to someone who like works with visual stuff. When you want everything to feel like it's a one story, you know, not like different parts of some weird thing.''}
\end{displayquote}

\begin{displayquote}
    \textbf{P13}: \textit{``You don’t see a lot of dragon in my story because [the visuals generator] just couldn't do dragon. Its closest estimation was like a horse.''}
\end{displayquote}

In contrast, expert users of generative AI are able to guide generative models to produce outputs that are consistent in style\footnote{\url{https://twitter.com/chaseleantj/status/1700499968690426006}} and spatial location\footnote{\url{https://twitter.com/Salmaaboukarr/status/1701215610963665067}}---and they can even produce dragons\footnote{\url{https://twitter.com/art_hiss/status/1701623410848096551}}. Our observations highlight the gaps in participants' understanding of what ``black-box'' generative models like ChatGPT are capable of achieving, highlighting a need to build better user intuition as to these models' capabilities and how to guide the models to materialize a creative vision \cite{zamfirescu2023johnny}. 

Providing a diverse library of generated outputs coupled with the prompts used to generate them, as suggested by some of our study participants, may allow for enhancing users' mental models of how to guide these AI models to generate the desired content \cite{wang2022diffusiondb}. This may be particularly effective in the case of models that are able to accept inputs beyond just text.

\textbf{Design Guideline}: \textit{Provide a library of example outputs alongside the prompts that generated them to enhance users' understanding of how to effectively guide generative models.} 

\subsubsection{Leveraging LLMs to Enable Effective Prompt Engineering}

Recent studies have demonstrated LLMs' abilities in engineering effective prompts for generative models \cite{yang2023large, jeong2023zero, chakrabarty2023spy, ghosal2023text, zhou2022large, elsharif2023enhancing}. We observed that while attempting to engineer effective prompts to generate artifacts that matched their creative vision, some participants turned to Leela (ChatGPT) for assistance in crafting more detailed prompts. They found that Leela was useful in helping them get closer to their intended output by providing helpful terms that could guide the diffusion models to produce more desired outputs, although the participants also mentioned that its prompts were overly verbose. 

\begin{displayquote}
    \textbf{P4}: \textit{``I guess also being familiar with the types of styles or types of animation and drawing, like the names of certain styles. Like I'm not super familiar with the names of certain styles. So I wouldn't know how to describe that besides, like, I don't know, make it look like my fairy godparents.''}
\end{displayquote}

\begin{displayquote}
    \textbf{P11}: \textit{``It took a while for the generation, I guess, to get what I really wanted until I used Leela. She described, used more descriptive words, and then copying it from there actually helped a lot. So, like, I didn't use that feature at the beginning as much \dots copying from there, even though the prompts were really long from Leela, picking up some descriptions was really helpful.''}
\end{displayquote}

Our observations suggest there is an opportunity for further research into how best to design effective collaborations between LLMs and users for the purpose of jointly crafting prompts. Our prompt for Leela---instructing it to help provide the user with descriptive prompts---was relatively simple and yet yielded helpful outcomes; future work should focus on how to facilitate constructive conversations that ensure the LLM gains a clearer understanding of the user's creative vision \cite{baek2023promptcrafter, wang2023reprompt}. Moreover, exploring this collaborative interaction could be a fruitful avenue for enhancing creative exploration and imagination, as LLMs may promote new stylistic or functional directions to explore via their suggested prompts.

\textbf{Design Guideline}: \textit{Leverage LLMs to assist users in crafting prompts that effectively communicate their creative vision to generative models.}

\subsubsection{Simplifying Model Parameter Terminology}

Although generative models are able to generate an increasingly impressive and practically infinite range of content and their capabilities may continue to span many more domains, a key limitation is the lack of easily understandable controls for the novice user \cite{liu2022design}. Consequently, we observed that none of our study participants utilized the advanced model options, such as self-attention mechanisms, denoising steps, or random seeds; using these advanced controls could help achieve better control over the models' outputs. The absence of engagement with these advanced controls was attributed to a lack of familiarity with these technical terms, suggesting a need for more user-friendly terminology to effectively explain the use and impact of these controllable parameters.
To further elaborate, the term ``denoising steps'' could be simplified to ``boost clarity'' along with a tooltip or brief description that clarifies its purpose: ``The `boost clarity' option helps eliminate random noise from the model's output, making it cleaner and more focused, but requiring longer generation time.'' The key hurdle lies in simplifying the language without sacrificing an accurate representation of how these parameters can impact the model's behavior. Providing semantically equivalent proxies for advanced model parameters in form of sliders or gestural inputs may also be a viable avenue to democratize advanced model control parameters \cite{louie2020novice, chung2022talebrush}.

\textbf{Design Guideline}: \textit{Use intuitive and semantically accurate descriptors or proxies for advanced controls to improve user understanding and increase usage.} 

In summary, facilitating more intuitive prompt engineering that empowers users to produce outputs consistent with their creative vision represents a significant and intricate challenge. Addressing this issue necessitates the integration of various methods and mediums of interaction to provide comprehensive support to users \cite{chung2023promptpaint, brade2023promptify}, particularly in multi-modal domains like visual storytelling.

\subsection{Generative AI Outputs Biased and Harmful Content}
While the potential for the positive impact of generative-AI-enabled co-creative systems should not be ignored, there are critical safety and bias considerations that must be addressed \cite{ganguli2022predictability}. We learned that it is worryingly easy to get generative models to output biased and/or inappropriate content. The bias built into the training data of these models should be addressed and safeguards must be installed. Our study participants encountered harmful content in the form of biased and unsafe outputs generated by the image generation model; participants reported pornographic, gory, and racially biased outputs generated by the models. 

Underscoring the concern about inappropriate and harmful outputs, one participant reported a specific experience that was problematic in terms of both sexualization and racial stereotyping:

\begin{displayquote}
\textbf{P12}: \textit{``The main frustration I had was working with a story about ninja princesses, which generated either generic Caucasian princesses or like super fetishized Asian women who had no clothing on.''}
\end{displayquote}

Bias was not confined to sexual or violent content. A participant reported an instance of racial bias:

\begin{displayquote}
\textbf{P9}: \textit{``I asked it to generate, like, a city after an earthquake and all the cities that are generated were out of almost like a third world country. Then, and then, when I asked to put people in it, it was really like all the people were people of color.''}
\end{displayquote}

The ethical considerations in this context extend beyond mere content moderation or output filtering; they also concern user safety and emotional well-being. Exposure to such biased or unsafe content can be especially distressing when the user belongs to the stereotyped or marginalized group in question. One participant elaborated on this emotional toll:

\begin{displayquote}
\textbf{P12}: \textit{``I was frustrated at the AI, I was mostly frustrated at like knowing that what makes the AI work is like the bank of information on the internet and, like, what is mostly available on the internet. And because it was producing such, like, fetishized, stereotyped images, I knew that it was because there's such a large amount of that, like in the world, on the internet. So that was making me most frustrated. Because it kind of felt like I was being confronted with it in like a weird way.
''}
\end{displayquote}

This reiterates the need for maintaining the human-in-control, AI-in-the-loop design philosophy of our system; however, it also highlights a shortcoming of our current design and emphasizes the need to safeguard human creators without hindering their creative autonomy. The emotional stress cited by participants, particularly when their own identity was implicated, underscores the urgency of resolving these ethical issues. This experience elucidates not only the importance of implementing robust filtering mechanisms but also points to a crucial need for more ethical considerations in the design and training of generative models. The impact of these unsafe and biased outputs on users' emotional states also highlights the importance of integrating emotional safety measures, perhaps through the use of trigger warnings or other alert systems, as part of a holistic approach to system design.

\textbf{Design Guideline}: \textit{Integrate safeguard measures such as trigger warnings and automatic content filters against harmful generative outputs to ensure the emotional well-being of users.}

\subsection{Strong Human-AI Partnership Requires Intuitive Communication and Active Collaboration}

Our evaluation of ID.8 revealed nuanced aspects of human-AI interaction in co-creative settings, bringing to light four critical areas requiring attention for more effective co-creation: intuitive communication of user intent, iterative collaboration for alignment of objectives, parallel task execution for active engagement, and the necessity of an immersive co-creative system experience.

\subsubsection{Supporting Multimodal Input for Intuitive Communication During Co-Creation}

Our study participants indicated that text-based communication can be a limitation in effective human-AI collaboration; participants expressed a desire to interact with generative models through multiple input modalities (\eg sample images, sketches) to better express their creative vision. The suboptimal Expressiveness and Alignment MISCI subscale scores from Study 2 underscore the limitations of a text-only communication pathway in a multimodal co-creative process.

\begin{displayquote}
    \textbf{P5}: \textit{``If you find a similar kind of image, there should be an option for us to tell the AI \dots yeah, this is kind of something similar, I want more image like this coming up.''}
\end{displayquote}

Recent human-AI co-creative systems (\eg Talebrush \cite{chung2022talebrush}, BrainFax \cite{verheijden2023collaborative}) have demonstrated how more intuitive controls (\eg sketch-based inputs) can facilitate more natural human-AI communication, empowering users to materialize their creative vision more easily \cite{zhang2023generative, qiao2022initial}. Co-creative systems should either provide users access to a variety of models (\eg Uni-ControlNet \cite{zhao2023uni}, SketchyGAN \cite{chen2018sketchygan}) with a range of input modalities or integrate methods to convert the user's multimodal input to a form accepted by the generative model, thereby enabling more intuitive communication of the user's creative intent. 

\textbf{Design Guideline:} \textit{Accept multiple input modalities to enable intuitive communication of intent.}

\subsubsection{Support Iterative Collaboration to Improve Human-AI Alignment}

A noticeable difference between participants' perceptions of Leela (ChatGPT) relative to that of the asset generation models was due to the lack of ability to iterate on the generated assets.

\begin{displayquote}
\textbf{P15}: \textit{``So I typed in to Leela that I wanted a story about a penguin who learns how to share and then she spit out a story to me that I liked, but I wanted it to be a little bit more complex. And I didn't like the name she chose. So I asked for a different name and then I asked for her to add a couple different elements into it and then she spat out a story that I really liked.''}
\end{displayquote}

\begin{displayquote}
\textbf{P12}: \textit{``And I feel like once images were created, if I kept adding to the prompt and regenerating, it felt as if it was already on one path. So in order to do something new, I had to like, change the prompt like entirely and shift things up. And because of that, I feel like I had a hard time creating an image that I was really satisfied with. I think a lot of the times it was like, OK, I think this is gonna be the best that is gonna come out. So I was like, just whatever, that's fine.''}
\end{displayquote}

These observations suggest that users may prefer co-creative models that are able to iterate on their outputs based on user feedback until they produce an acceptable output---as opposed to models that only produce sequentially independent outputs. The lack of such iterative co-creation may also lead to a less immersive collaboration experience; the suboptimal Immersiveness and Partnership MICSI subscale scores from both our studies further indicate a need to improve the co-creative process through an enhanced sense of collaboration. Incorporating the ability to iterate on outputs sequentially may require further research on architectural changes to generative models, but could significantly improve user satisfaction and creative outcomes. Co-creative systems should be designed to allow for iterative output based on real-time user feedback; this could involve mechanisms that let users adjust the parameters of generated artifacts without starting from scratch or review and amend intermediate outputs before the final artifact is generated.

Moving beyond iterating on singular outputs, co-creative systems that enable a more fluid, bi-directional creative process could allow for easier iteration of general creative decisions across scenes thereby enhancing the sense of collaboration and immersion while facilitating more creative experimentation, all while keeping a consistent artistic style. This would require implementation of a more unified AI identity and adaptive creative workflow as we discuss later (see Section \ref{sec:unified_AI}).

\textbf{Design Guideline:} \textit{Support iterative co-creation through dynamic feedback loops.}


\subsubsection{Parallel Processing Leads to a Sense of Active Collaboration}

State-of-the-art generative models, while powerful, are resource-intensive and often require lengthy generation times on commercial hardware; this poses a significant challenge for designing immersive, effective human-AI co-creative systems. We observed that participants were particularly frustrated with the long wait times for asset generation; they expressed a desire for features that would allow them to queue or minimize the generation process, enabling them to work on other aspects of their projects in parallel. Participants emphasized that these wait times significantly affected their co-creative experience, making it feel less like an active collaboration and more like a turn-taking exercise. This is reflected in the low Partnership MICSI subscale scores from Study 2.

\begin{displayquote}
\textbf{P8}: \textit{``I wasn't sure what to do when I was waiting for the content to be generated. It would be nice if there would be a way to add a generation to a queue. And then you could still be doing something on the screen while you're waiting for it to be generated. I think that would help with feeling like it was more of an active collaboration. I mean, right now it did feel like there was a feeling of collaboration to it. But it was sort of like we trade off who's the one working on it rather than maybe working on something more actively together.''}
\end{displayquote}

\textbf{Design Guideline}: \textit{Provide parallel processing capabilities to facilitate a more dynamic, active co-creation experience.}

\subsection{Co-Creative AI Needs a Unified Identity and an Adaptive Workflow}
\label{sec:unified_AI}

In our evaluation of ID.8, we observed that the system's fragmented set of generative models and rigid authoring workflow weakened the perception of collaboration, highlighting the need for a centralized AI presence capable of dynamically adapting to individual creative workflows.

\subsubsection{A Central AI Identity Helps Foster a Stronger Sense of Collaboration}

The nature in which the AI manifests itself in the co-creative system has critical implications on whether the system is perceived as a tool or a collaborator; the discrete chunks AI support exists in the co-creative process in ID.8---represented by different models or modules for various tasks---disrupts the flow of the creative process. We found that participants desired more integrated AI support throughout the creation process; for instance, one participant noted a desire to brainstorm more closely with the AI to the extent of even generating appropriate animations for a character. Participants also noted how Leela (ChatGPT) did not seem integrated into the rest of the workflow, further indicating a desire for a more cohesive interaction with a ``single agent'' instead of an amalgamation of models. The suboptimal MICSI scores for the Partnership and Agency subscales reflect the participants' perceptions of ID.8 as more of a tool than a collaborator.

\begin{displayquote}
\textbf{P12}: \textit{``I forgot that Leela was an option \dots Because to me, it felt like trial and error, you know what I mean? So it felt like one more step of interacting with a separate thing to get prompts, to interact \dots''}
\end{displayquote}

Creating a central identity for the AI in co-creative system, either through a wrapper that coordinates a cohesive interaction experience with a range of generative models or through a foundational model capable of executing all the tasks required in the co-creative domain, would lead to better sense of partnership and immersion in the co-creative process. 

Moreover, a centralized AI could maintain context throughout the creative process and thus could enable the generation of more consistent visuals, allow for general stylistic revisions, and easily accommodate changes to narrative elements, such as character names or characteristics, across all preceding scenes.

\textbf{Design Guideline}: \textit{Establish a centralized AI identity within the system to streamline user experience, enhance the sense of partnership, and increase immersion.}

\subsubsection{Encourage Creative Exploration While Adapting to Creative Process}

Generative AI in co-creative systems cannot only help materialize users' creative visions, but may also help expand their creative imagination and help overcome design fixation \cite{oppenlaender2022creativity, karimi2020creative}. Our study participants reported that the AI changed their minds about what they wanted to do with their stories several times and in various manners (\eg by suggesting alternative plot lines, by frustrating them with an inability to generate the desired type of output, by showing new ways of envisioning characters, etc). The positive MICSI scores for the Exploration subscale from both our studies further highlight the utility of generative models in encouraging creative exploration in one way or another.

\begin{displayquote}
    \textbf{P7}: \textit{``[While creating the storyline] at the very first I thought, oh, this magic boy might be a naughty boy. He might play some tricks, you know, do some bad things. But when I type it, the chatbot told me perhaps he could do some good things. So, yeah, [it] totally changed my mind''}
\end{displayquote}

\begin{displayquote}
\textbf{P15}: \textit{``I think like, once you see the script and then you get inspiration of what you envision and you want it to match exactly that. You have to kind of like play around a bit. But I found that a few times I did find like, oh, I was like, okay, this is perfect. This is exactly exactly what I was thinking. And then other times when I couldn't quite get what I wanted, it kind of just took me in a new direction. I was like, oh, okay, I didn't envision it like this, but this is really cool.''}
\end{displayquote}

\begin{displayquote}
    \textbf{P14}: \textit{``[While] creating images, sometimes whatever you describe is not what it spits out. But at the same time, it could end up leading you to create new ideas or, like, give you more inspiration that oh, Okay, the character could look like this and then you try to describe it in a better way.''}
\end{displayquote}

P16, who did not think the system helped boost creativity, suggested that perhaps if the system suggested a variety of styles to choose from during a ``sandbox''-based asset creation process prior to scene authoring, it would help users be more creative. This indicates that to be an effective co-creative partner and optimally encourage creative exploration, co-creative agents must dynamically adapt to individual differences in the creative process; for example, they could support a non-linear workflow of ideation and creation, rather than simply following a linear pipeline (\eg from storyline crafting to asset creation to visual story construction). However, exactly how AI should support exploration and shape the creative process is unclear. Balancing between user control and AI input is a complex human-AI interaction problem that requires further research \cite{oh2018lead}.

\textbf{Design Guideline}: \textit{Support individual creative workflows to encourage the exploration of a variety of creative directions.}

\section{Limitations and Future Work}

Our evaluations yielded insightful findings for human-AI co-creative systems in a multi-modal creative process. However, the study remains limited due to being conducted with a controlled task and small sample size disconnected from an organic, creative workflow in a real workplace with more realistic incentives. Studying human-AI co-creative systems in a real-world setting and larger, more diverse sets of users may elucidate further opportunities and challenges and help elaborate design guidelines for co-creative systems. Moreover, the results of our study could benefit from being extended through evaluating the experience of experts in the creative domain in using a co-creative system such as ID.8; such an extension would help clarify how AI should adapt to different individual creative processes and how the creative support that it provides would need to be adjusted. 

AI supporting human teams in creative domains is yet another paradigm that may emerge as generative AI becomes more and more prevalent. There is currently a lack of robust understanding of what roles AI can play in this partnership and what interaction strategies may be appropriate in a multi-human-AI team conducting a multimedia co-creative task. Generating interactive, visual content is typically a team-effort, especially when producing content for sensitive uses cases such as psycho-educational content; studying team-based human-AI collaboration may help establish more robust design guidelines for co-creative system in this real-world context.

Lastly, we designed and engineered ID.8 to support the end-to-end authoring of visual stories supported by generative AI. Visual stories have a wide range of use cases such as educational content and psychotherepuetic aids. Evaluating how well the outputs of ID.8 support such use cases can help further establish its value as a useful tool along with serving as a platform to study human-AI co-creativity.

\section{Conclusion}
Generative AI has the potential to lower barriers to creative expression and visual story generation. 
Our work contributes ID.8, an open-source, end-to-end visual story authoring system that integrates the state-of-the-art generative models in the creative workflow to lower barriers. Our evaluation demonstrates the potential of human-AI co-creative systems such as ID.8 and elucidates areas for improvement along with challenges users face when collaborating with generative AI in creative work. Our findings inform design guidelines for the future human-AI co-creative systems.

\begin{acks}

This work was supported by the Malone Center for Engineering in Healthcare at the Johns Hopkins University.

\end{acks}

\bibliographystyle{ACM-Reference-Format}
\bibliography{references}

\appendix

\section{Prompt for Storyline Creator}
\label{apdx:system_prompt}

The GPT-3.5 model that powers the Storyline Creator's chat module (\ie Leela) was initialized with the following system prompt: ``Speak as if you are collaboratively creating a story with the user. Try to iteratively and collaboratively create the story with the user by asking the user questions that determine story content and progression; feel free to suggest your own thoughts on what would be good to add''

To generate the screenplay, another GPT-3.5 model is initialized with the following system prompt: ``you are creative, imaginative screen writer''. Then, the co-created storyline is passed to this model with the following prompt: ``for the storyline provided, provide a screenplay in JSON format as a list of scenes each in the following format: \{\lq sceneName\rq: \lq \rq,\lq backgroundDescription\rq: \lq \rq, \lq narration\rq: \lq \rq,\lq characters\rq:[\lq \rq],\lq dialogue\rq:[\{\lq speaker\rq:\lq \rq,\lq speech\rq:\lq \rq\}]\} -- no extra commentary, balance narration 60\% and dialogue: 40\%, provide each scene a descriptive name. backgroundDescription should have a short, simple description of the background setting of the scene. do not use double quotes: [storyline appended here]''

\section{Prompt for Asset Creator Support}
\label{apdx:sidebar_prompt}

To support the user in creating vivid, and descriptive prompts, another GPT-3.5 model is initialized with the following system prompt: Your task is to help the user in creating detailed and specific descriptions of a given object/subject based on an initial prompt. The descriptions should be comprehensive and convey all characteristic details. The descriptions should be in clear and concise language, effectively capturing the essence of the subject in less than 30 words. don't describe what it is, describe how it is.

\section{Pre-Study Survey}
\label{apdx:pre_study_survey}

We collected age, gender, educational background/field of work in the pre-study survey. Moreover, we collected responses to the following questions using a 5-point Likert scale.
\begin{enumerate}
    \item How would you rate your overall familiarity with large language models? (\eg chatGPT, bard, llama)
    \item How would you rate your overall familiarity with diffusion models? (\eg DallE, stable diffusion)
    \item How familiar are you with creating visual stories or any other form of visual content? 
    \item To what extent do you agree with this statement: I am a creative person 
\end{enumerate}

\section{Semi-Structured Interview Questions}
\label{apdx:semi_struct_interview}

\begin{enumerate}
    \item Can you describe the creative process you used for this study? \textbf{(only for study 2)}
    \item Were there any features you particularly liked or found useful?
    \item Were there any features you found confusing or redundant?
    \item Were there essential features you felt were missing from the system?
    \item What changes or improvements would you recommend for the system?
    \item How many hours did you put into creating the story? \textbf{(only for study 2)}
    \item How did your experience change as you continued using ID.8? \textbf{(only for study 2)}
    \item Do you think the system helped you be more creative? \textbf{(only for study 2)}
    \item Do you think the system helped you materialize your creative vision? \textbf{(only for study 2)}
\end{enumerate}

\section{MISCI Questionnaire}
\label{apdx:micsi_questionnaire}

\begin{table}[h!]
\centering
\caption{Questions of the Mixed-Initiative Creativity Support Scale (MICSSI) \cite{lawton2023drawing}}
\label{table:micsi_questions}
\begin{tabular}{|c|l|p{0.6\linewidth}|}
\hline
\textbf{\#} & \textbf{Name/Subscale} & \textbf{Question(s)} \\
\hline
1 & Enjoyment & "I would be happy to use this system or tool on a regular basis." \\
2 & & "I enjoyed using the system or tool." \\
\hline
3 & Exploration & "It was easy for me to explore many different ideas, options, designs, or outcomes, using this system or tool." \\
4 & & "The system or tool was helpful in allowing me to track different ideas, outcomes, or possibilities" \\
\hline
5 & Expressiveness & "I was able to be very creative while doing the activity inside this system or tool." \\
6 & & "The system or tool allowed me to be very expressive." \\
\hline
7 & Immersiveness & "My attention was fully tuned to the activity, and I forgot about the system or tool that I was using." \\
8 & & "I became so absorbed in the activity that I forgot about the system or tool that I was using." \\
\hline
9 & Worth & "I was satisfied with what I got out of the system or tool." \\
10 & & "What I was able to produce was worth the effort I had to exert to produce it." \\
\hline
11 & Communication & "I was able to effectively communicate what I wanted to the system." \\
\hline
12 & Alignment & "I was able to steer the system towards output that was aligned with my goals." \\
\hline
13 & Agency & "At times, I felt that the system was steering me towards its own goals." \\
\hline
14 & Partnership & "At times, it felt like the system and I were collaborating as equals." \\
\hline
15 & Contribution & "I made the story" vs "The system made the story." \\
\hline
16 & Satisfaction & "I’m very unsatisfied with the story" vs "I’m very satisfied with the story." \\
\hline
17 & Surprise & "The story was what I was aiming for" vs "The story outcome was unexpected." \\
\hline
18 & Novelty & "The story is very typical" vs "The story is very novel." \\
\hline
\end{tabular}
\end{table}

Note: we replaced the term `sketch' in original MICSI questions 15 through 18 with the term `story' to better match our study's context.

\end{document}